\documentclass[journal]{IEEEtran}
\usepackage{caption}
\usepackage{amsmath,amsfonts}
\usepackage{algorithmic}
\usepackage{algorithm}
\usepackage{array}
\usepackage{textcomp}
\usepackage{stfloats}
\usepackage{url}
\usepackage{verbatim}
\usepackage{graphicx}
\usepackage{cite}
\usepackage{color}
\usepackage{amsmath}
\usepackage{times}
\usepackage{epsfig}
\usepackage{amssymb}
\usepackage{booktabs}
\usepackage{bm}
\usepackage{pifont}
\usepackage{amsthm}
 
\usepackage{subfigure}

\usepackage[table,xcdraw]{xcolor}
\usepackage{soul}

\usepackage{tikz}
\usepackage{hyperref}
\usepackage{orcidlink}
 
\usepackage{ragged2e} 
\usepackage{booktabs,makecell, multirow, tabularx}
 
\hypersetup{
colorlinks=False,
linkcolor=red,
citecolor=green
}
 
\def\ie{$i.e.$}
\def\eg{$e.g.$}
\def\etc{$etc. $}

\usepackage{amsthm}
\usepackage{arydshln}

\def\D{\mathcal{D}}

\def\X{\mathcal{X}}
\def\Y{\mathcal{Y}}

\usepackage{enumitem} 

\def\btheta{\boldsymbol{\bm{\theta}}}

\def\x{\bm{x}}

\begin{document}

% \title{A Sample Article Using IEEEtran.cls\\ for IEEE Journals and Transactions}
\title{WPDA: Frequency-based Backdoor Attack with Wavelet Packet Decomposition}
\author{ Zhengyao Song, Yongqiang Li, Danni Yuan, Li Liu, Shaokui Wei and Baoyuan Wu
% IEEE Publication Technology,~\IEEEmembership{Staff,~IEEE,}
\thanks{Zhengyao Song and Yongqiang Li are with School of Instrumentation Science and Engineering, Harbin Institute of Technology, Harbin, 150000, Heilongjiang Province, China,
% School of Instrumentation Science and Engineering, Harbin Institute of Technology, Harbin, China, 
e-mail: songzhengyao@stu.hit.edu.cn, liyongqiang@hit.edu.cn.}

\thanks{Danni Yuan, Shaokui Wei and Baoyuan Wu are with School of Data Science, The Chinese University of Hong Kong, Shenzhen, Guangdong, 518172, P.R. China, 
% the School of Data Science,
% The Chinese University of Hong Kong, Shenzhen, China, 
email: danniyuan@link.cuhk.edu.cn, shaokuiwei@link.cuhk.edu.cn, wubaoyuan@cuhk.edu.cn.
} 
\thanks{Li Liu is with Artificial Intelligence Thrust, Information Hub, Hong Kong University of Science and Technology (Guangzhou), Guangzhou, 511455, Guangdong Province, China,
% the Hong Kong Universityof Science and Technology (Guangzhou), China, 
email: avrillliu@hkust-gz.edu.cn.}
\thanks{Corresponding Author(s): Yongqiang Li and Baoyuan Wu.
}}

% \hspace{0.5mm}$^{~\orcidlink{0000-0001-1234-1234}}$
%here
\markboth{}%IEEE Transactions on Information Forensics and Security
{Shell \MakeLowercase{\textit{et al.}}: A Sample Article Using IEEEtran.cls for IEEE Journals}

% \IEEEpubidadjcol
\maketitle
\begin{abstract}

This work explores backdoor attack, which is an emerging security threat against deep neural networks (DNNs). The adversary aims to inject a backdoor into the model by manipulating a portion of training samples, such that the backdoor could be activated by a particular trigger to make a target prediction at inference. 
Currently, existing backdoor attacks often require moderate or high poisoning ratios to achieve the desired attack performance, but making them susceptible to some advanced backdoor defenses (\eg, poisoned sample detection). 
One possible solution to this dilemma is enhancing the attack performance at low poisoning ratios, which has been rarely studied due to its high challenge. To achieve this goal, we propose an innovative frequency-based backdoor attack via wavelet packet decomposition (WPD), which could finely decompose the original image into multiple sub-spectrograms with semantic information. It facilitates us to accurately identify the most critical frequency regions to effectively insert the trigger into the victim image, such that the trigger information could be sufficiently learned to form the backdoor. The proposed attack stands out for its exceptional effectiveness, stealthiness, and resistance at an extremely low poisoning ratio. Notably, it achieves the $98.12\%$ attack success rate on CIFAR-10 with an extremely low poisoning ratio of $0.004\%$ (\textit{i.e.}, only 2 poisoned samples among 50,000 training samples), and bypasses several advanced backdoor defenses. 
Besides, we provide more extensive experiments to demonstrate the efficacy of the proposed method, as well as in-depth analyses to explain its underlying mechanism. 

% This work explores backdoor attack, which is an emerging security threat against deep neural networks (DNNs). The adversary aims to inject a backdoor into the model by manipulating a portion of training samples, such that the backdoor could be activated by a particular trigger to make a target prediction at inference. 
% Currently, existing backdoor attacks often require high poisoning ratios to achieve the desired attack performance, making them susceptible to detection and defense techniques. Therefore, the exploration of backdoor attacks at low poisoning ratios is challenging and significant. To achieve this goal, we propose a novel frequency-based backdoor attack via wavelet packet decomposition (WPD), an advanced signal decomposition technique that finely decomposes the original image signal into multiple spectrograms with semantic information, allowing us to accurately find out the most critical frequency regions to achieve precise poisoning. Our method stands out for its exceptional stealthiness, effectiveness and stability. Notably, it achieves the $98.12\%$ Attack Success Rate (ASR) on CIFAR-10 with an extremely low poisoning ratio of $0.004\%$ (\textit{i.e.}, only 2 poisoned samples among 50,000 training samples) and bypassing most existing detection and defense methods. Besides, we also provide comprehensive experiments and analyses to explain the underlying mechanism.

\end{abstract}
\begin{IEEEkeywords}
Backdoor attack, low poisoning ratios, wavelet packet transform
\end{IEEEkeywords}
\section{Introduction}
\normalsize
\IEEEPARstart{D}{eep}
neural networks (DNNs) have achieved remarkable success and widespread adoption in various research fields (\eg, face recognition~\cite{balaban2015deep}, voice recognition~\cite{li2012improving,yuan2018commandersong}, and autonomous driving~\cite{zhou2020deepbillboard}), leading to significant advancements in artificial intelligence. However, this progress also introduces potential security vulnerabilities. 
Among these security threats, backdoor attacks pose a significant risk, involving maliciously manipulating a DNN model during training to embed hidden backdoors. These backdoors can be triggered by specific input patterns, causing the model to produce targeted incorrect outputs while maintaining normal performance on benign inputs. The implications of such attacks range from targeted misclassification and data manipulation to compromising the integrity and reliability of critical systems that rely on DNNs.

In this work, we study the threat model of data poisoning-based backdoor attack, where the attacker can only manipulate the training dataset, without access to the training process. It applies to the practical scenario where the user doesn't have sufficient training data and thus downloads or buys data from an untrustworthy third-party data platform. Although several effective backdoor attacks have been developed, they often require a substantial number of poisoned training samples to achieve the desired attack success rate, while potentially increasing the risk of suspicion to backdoor defenses. 
One possible solution is to implement effective attack at low poisoning ratios, which is still very challenging until now (please refer to the evaluations presented in \cite{wu2022backdoorbench}). 
%In scenarios of backdoor attacks with a low poisoning ratio, the primary challenge is ensuring that the model can effectively learn the trigger from a limited number of poisoned training samples. 
We believe the key to achieve that goal is that \textit{the trigger information could be sufficiently learned by the model from the limited number of poisoned samples}. 

Inspired by previous studies~\cite{wang2020high,ilyas2019adversarial,rahaman2019spectral,abello2021dissecting,wang2023neural} which reveal that DNNs can efficiently learn some specific frequency signals of images, we focus on identifying critical regions in the frequency domain to design backdoor triggers. 
Specifically, we adopt the Wavelet Packet Decomposition (WPD)~\cite{xiong1998wavelet}, which provides a fine-grained multi-scale frequency decomposition, effectively capturing frequency signals of images at various scales. 
Based on this decomposition, we further present an in-depth analysis about the learning mechanism of DNNs in frequency domain, and design an effective approach to automatically identify the most critical frequency regions that can be sufficiently learned by the DNN model for each dataset. Moreover, we propose an elaborately designed strategy to insert the backdoor trigger information into the identified critical frequency regions in the training and testing dataset, to ensure that the trigger information can be sufficiently learning during training, and can effectively activate the backdoor during testing while maintaining visual stealth. 
Our evaluations shows that, with an extremely low poisoning ratio $0.004\%$, the attack success rate (ASR) of the proposed attack method could achieve $98.12\%, 85.01\%, 80.99\%$ on CIFAR-10, CIFAR-100, Tiny ImageNet, respectively. In contrast, the ASRs of all compared attacks are lower than $13\%$ with the same low poisoning ratio. 
Moreover, we also show that the proposed attack can effectively bypass several advanced backdoor defenses, and present in-depth analysis to reveal the underlying mechanism of our method. 
% To the best of our knowledge, it is the first time to achieve such high attack performance with 

The main contributions of this work are threefold. 
\begin{itemize}
\item We leverage WPD to demonstrate the learning mechanism of DNNs in the frequency domain, and propose an effective approach to identify the most critical frequency regions that could be sufficiently learned by the DNN model.  
\item We develop an innovative backdoor attack method according to the identified critical frequency regions, referred to as WPDA, which is characterized by its effectiveness, stealthiness, and resistance with extremely low poisoning ratio. 
\item We provide extensive experiments to verify the efficacy of the proposed method, as well as in-depth analyses to explain its underlying mechanism.
\end{itemize}

\section{Related Work}
\label{related work}
\subsection{Backdoor Attack}
Current research about backdoor attacks can be interpreted as the process of maliciously implanting a backdoor into the model, which would induce the model to build the relationship between special input triggers with a desired output. The main methodologies include poisoning training data~\cite{gu2019badnets,chen2017targeted,li2021invisible,shafahi2018poison,gao2023imperceptible}, manipulating the model's parameters~\cite{kurita2020weight
% ,9737144
}, or controlling the training process~\cite{
shumailov2021manipulating, 
nguyen2021wanet,nguyen2020input,bagdasaryan2021blind,jiang2022incremental}. In particular, attackers can manipulate both images and labels~\cite{gu2019badnets,chen2017targeted,li2021invisible} in data poisoning works.

% there are two types of attack settings, in one of which attackers can manipulate both images and labels~\cite{gu2019badnets,chen2017targeted,li2021invisible}, while in the other setting that attackers can only modify labels~\cite{shafahi2018poison,zhao2020clean}. 
BadNets~\cite{gu2019badnets} is a pioneering work on backdoor attacks in image classification tasks that inserts a white patch into benign samples. Subsequent research has explored various backdoor attack techniques, including Blended~\cite{chen2017targeted}, which takes a cartoon image as the trigger and blends it with the benign samples. Liu et al.~\cite{liu2020reflection} propose a reflection backdoor attack,  which enhances backdoor attack by applying random spatial transformations to the poisoned image. While such methods have demonstrated effectiveness, the triggers often introduce noticeable visual artifacts, making them susceptible to backdoor defenses.
To improve the stealthiness of backdoor attacks, some dynamic invisible trigger generation methods have been proposed, including sample-specific invisible triggers generated by the DNN-based image steganography~\cite{li2021invisible,li2020invisible}, autoencoder neural network~\cite{ning2021invisible,zhang2022poison}, random noise optimizations by adversarial perturbation~\cite{wei2022adversarial,zhang2021advdoor,wang2021psat,che2021adversarial}, image warping~\cite{nguyen2021wanet,cheng2020deep} or RGB filter~\cite{gong2023kaleidoscope}, \etc~
\normalsize

Instead of inserting triggers in the spatial domain, Wang et al.~\cite{wang2021backdoor} indicated that small perturbations in the frequency domain could produce perturbations dispersed across the entire image, and they proposed a backdoor attack called {\small{FTROJAN}} which inserts triggers in mid- and/or high-frequency components of UV channel by Discrete Cosine Transform (DCT). Besides, Feng et al.~\cite{feng2022fiba} embed a trigger into the amplitude spectrum while maintaining the phase spectrum unchanged by utilizing Discrete Fourier Transform (DFT)~\cite{brigham1988fast}. ~\cite{Hammoud_2022_BMVC} generates a poisoning filter based on a Fourier heatmap to poison part of the training samples. Drager et al.~\cite{drager2022backdoor} propose a wavelet transform-based attack (WABA) method which inserts the trigger image into the lowest-frequency components of poisoned images by wavelet transform. These methods adopt a fixed selection strategy to select poisoning regions, without considering the difference among various datasets, resulting in suboptimal performance. To address this issue, we analyze the datasets by WPD to infer the critical frequency regions that the DNNs focus on and propose a dataset-specific backdoor attack method.

% \normalsize
% Unlike DFT and DCT, which focus primarily on global frequency representations, the wavelet transform offers the advantage of capturing both frequency and spatial information. This allows for the extraction of detailed features such as edges, contours, and textures, providing a more comprehensive understanding of image characteristics. Additionally, the multi-resolution analysis capabilities of the wavelet transform enable the decomposition of information at various scales, further enhancing the precision of feature extraction.
% Drager et al.~\cite{drager2022backdoor} propose a wavelet transform-based attack (WABA) method which inserts the trigger image into the lowest-frequency components of poisoned images by wavelet transform. 
% However, in most cases, it is difficult to maintain the naturalness of the poisoned images by inserting trigger information into the low-frequency region.
% Wavelet packet decomposition (WPD) is an improvement based on the wavelet transform, which is able to subdivide both high and low frequency information, while wavelet transform can only subdivide low frequency information. 
% Analysing the dataset by WPD to infer the critical frequency regions that the DNNs focus on is conducive to the proposal of an effective and precise backdoor attack method. In addition, how to design the trigger information to ensure the effectiveness of the backdoor attack while also making the poisoned images stealthy needs to be further explored.

\subsection{Backdoor Defense}
\normalsize
Backdoor defense can be broadly divided into three categories, which are pre-training, in-training, and post-training. For pre-training methods, the defender aims to detect poisoned samples in the training dataset by capturing the anomalous behavior characteristics of poisoned samples, such as activation of samples (\textit{e.g.}, AC\cite{chen2018detecting}, Beatrix\cite{mabeatrix}, SCAn\cite{tang2021demon}, Spectral\cite{tran2018spectral}, and SPECTRE\cite{hayase2021spectre}) and the prediction consistency of samples (\textit{e.g.}, CT\cite{qitowards} and ASSET\cite{asset}). For in-training methods, defenders distinguish poisoning samples and benign samples by the difference of their training loss \cite{li2021anti} or feature gathering \cite{huang2022backdoor, chen2022effective,zeng2021adversarial}, and then unlearn these samples or securely retrain a clean model on the processed dataset. For post-training methods, the defenders remove the embedded backdoors by snipping out possible backdoor neurons, including CLP~\cite{zheng2022data}, and NAD~\cite{li2021neural}, or fine-tuning the backdoored model to mitigate backdoor~\cite{wei2023shared,zhu2023neural,yao2024reverse}. 
Furthermore, defenders are capable of reversing backdoor triggers and then pruning backdoor models using these trigger-stamped samples, such as in the NC~\cite{wang2019neural} method, or by rejecting queries once the inputs are identified as poisoned samples (\textit{e.g.}, STRIP\cite{gao2019strip}
% , SCALE-UP\cite{SCALE-UP}, 
and TeCo\cite{liu2023detecting}).
\vspace{-0.5cm}
\begin{figure}[htbp]
\centering
\includegraphics[width=3in]{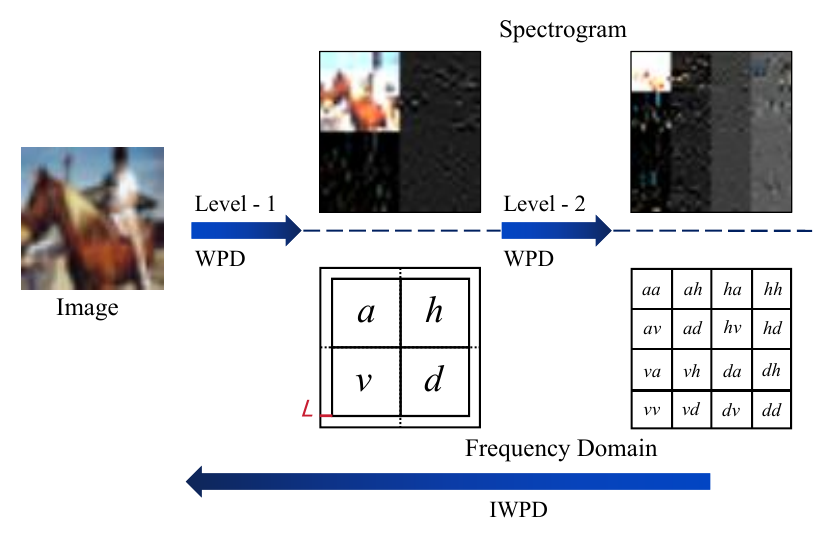}
\caption{Principle of wavelet packet decomposition. Region \textit{`a'} contains low frequency, region \textit{`h'} contains high-horizontal frequency, and region \textit{`v'} contains high-vertical frequency, Region \textit{`d'} contains high-diagonal frequency. WPD requires padding $L$ pixels on each edge of the image.}
\label{WPD}
\end{figure}
\vspace{-0.2cm}
\section{Method}
\label{Method}
\subsection{Preliminary}
Wavelet Packet Decomposition (WPD) can transform an image from the spatial domain to the frequency domain and decompose it into multiple frequency regions containing different semantic information. It
inherits the time-frequency analysis characteristics of the wavelet decomposition (WD). The main difference is that WD can only decompose low frequency, while WPD can decompose information in every frequency band. During the WPD process, the vanishing moments of the wavelet basis functions significantly impact the handling of image dimensions. Assuming that the vanishing moment is $2L$, it is necessary to pad $L$ pixels around each edge of the image to ensure that the image boundaries are effectively processed and to avoid information loss during decomposition. Fig.~\ref{WPD} illustrates the principle of WPD, assuming the shape of input RGB image $\x$ is $M \times M \times 3$, Eq.~\ref{WPD_equ} illustrates that after $N$-level WPD, $\x$ will produce $4^{N-1}$ parent-spectrograms, each containing 4 sub-spectrograms: 

\begin{equation}
\label{WPD_equ}
\begin{aligned}
\mathcal{W}^{N}(\x)
%  &= \sum_{m=0}^{M-1} \sum_{n=0}^{M-1} \sum_{c=0}^{2} \x(m, n, c) \cdot \psi(i - 2m) \cdot \psi(j - 2n) \cdot \psi(k - 2c) \\
% &
= \begin{bmatrix}
  C^N_{1,1}(\x)  & C^N_{1,2}(\x)  & \dots  & C^N_{2^{N-1},2}(\x) \\
  C^N_{1,3}(\x)  & C^N_{1,4}(\x)  & \dots  & C^N_{2^{N-1},4}(\x) \\
  \vdots & \vdots & C^N_{\textbf{p},\textbf{s}}(\x) & \vdots \\
  \vdots  & \vdots  & \dots  & C^N_{4^{N-1},4}(\x)
\end{bmatrix}.
% ,\\
 % \mathrm{s.t.}\quad & 0\le i=j \le {M+2L-1},~0\le m=n \le M-1.
\end{aligned}
\end{equation}
$C^N_{\textbf{p},\textbf{s}}(\x)$ represents the coefficient matrix of $\textbf{s}$-th sub-spectrogram in $\textbf{p}$-th parent-spectrogram with the shape of $\frac{M+2L}{2^N} \times \frac{M+2L}{2^N} \times 3$.
% The description of WPD for each channel $c$ of the input image $\x$ is in Eq.~\ref{WPD_equ3}:\\
Take 1-level WPD as an example, the coefficient matrices for the sub-spectrograms \textit{`a'}, \textit{`h'}, \textit{`v'} and \textit{`d'} are denoted as $C^1_{1,1}$, $C^1_{1,2}$, $C^1_{1,3}$ and $C^1_{1,4}$, respectively. 
\begin{equation}
\label{WPD_equ3}
\small
\begin{aligned}
% &C^1_{1,1}(i',j',c)= \sum_{c=0}^{2} \sum_{j=0}^{M-1} \sum_{j=0}^{M-1} \x(i, j, c) \cdot \psi_{p,s}(i' - 2i) \cdot \phi_{p,s}(j' - 2j),\\
&C^1_{1,1}(i',j',c)=  \sum_{i=1}^{M} \sum_{j=1}^{M} \x(i, j, c) \cdot h_{\overrightarrow{x}}(i' - 2i) \cdot h_{\overrightarrow{y}}(j' - 2j),\\
&C^1_{1,2}(i',j',c)=  \sum_{i=1}^{M} \sum_{j=1}^{M} \x(i, j, c) \cdot g_{\overrightarrow{x}}(i' - 2i) \cdot h_{\overrightarrow{y}}(j' - 2j),\\
&C^1_{1,3}(i',j',c)=  \sum_{i=1}^{M} \sum_{j=1}^{M} \x(i, j, c) \cdot h_{\overrightarrow{x}}(i' - 2i) \cdot g_{\overrightarrow{y}}(j' - 2j),\\
&C^1_{1,4}(i',j',c)=  \sum_{i=1}^{M} \sum_{j=1}^{M} \x(i, j, c) \cdot g_{\overrightarrow{x}}(i' - 2i) \cdot g_{\overrightarrow{y}}(j' - 2j),\\
% &\mathcal{W}^{N}_c(i,j)= \sum_{m=0}^{M-1} \sum_{n=0}^{M-1} \x(m, n, c) \cdot \psi_{p,s}(i - 2m) \cdot \phi_{p,s}(j - 2n),\\
&\mathrm{s.t.}\quad i'=j'\quad and\quad i',j' \in \{1,\dots,M+2L\},\\
&\quad \quad i=j\quad and\quad i,j \in \{1,\dots,M\},\quad c \in \{1,\dots,3\}.
 % &\mathrm{s.t.}\quad 1\le i=j \le M,~ 1\le i'=j' \le \frac{M+2L}{2^N} ,~ 1\le c \le 3.
\end{aligned}
\end{equation}

\noindent In Eq.~\ref{WPD_equ3}, $i'$ and $j'$ are used to describe the position of a pixel in the processed image, $i$ and $j$ are used to describe the position of a pixel in the input image. $c$ represents the color channel. $h_{\overrightarrow{x}}$ and $h_{\overrightarrow{y}}$ denote the low-pass filters in the horizontal and vertical directions, respectively. $g_{\overrightarrow{x}}$ and $g_{\overrightarrow{y}}$ denote the high-pass filters in the horizontal and vertical directions, respectively.
% $\psi_{p,s}$ and $\phi_{p,s}$ are wavelet packet basis functions that can be regarded as filters used for multi-scale and multi-band analysis in WPD. 
Note that $N \in \{1, 2, \ldots , \log_2^M\}$, $\textbf{s} \in \{1, 2, 3, 4 \}$, $\textbf{p} \in \{1, 2 ,\ldots, 4^{N-1} \}$, and the inverse transformation of the wavelet packet decomposition (IWPD) is denoted as $\mathcal W^{-N}(\x)$.
% The shape of the coefficient matrix $C^N_{p,s}$ of $s$-th sub-spectrogram in $p$-th parent-spectrogram is $\frac{M+2L}{2^N} \times \frac{M+2L}{2^N} \times 3$.
% \noindent The coefficient matrix of $s$-th sub-spectrogram in $p$-th parent-spectrogram can be obtained in Eq.~\ref{WPD_equ2}, where $(i,j)$ represents the position of a pixel in the processed image, $(m,n)$ represents the position of a pixel in the input image
% \begin{equation}
% \label{WPD_equ2}
% \begin{aligned}
%  \quad &C^N_{p,s}(i, j) = \sum_{m=0}^{M-1} \sum_{n=0}^{M-1} \x(m, n) \cdot \psi_{p,s}(i - 2m) \cdot \psi_{p,s}(j - 2n),\\
% \quad & \mathrm{s.t.}\quad 0\le i \le \frac{M+2L}{2^N},~0\le j \le \frac{M+2L}{2^N},\\
%  &\quad \quad ~0\le m \le M,~0\le n \le M.
% \end{aligned}
% \end{equation}

% The shape of the coefficient matrix of $s$-th sub-spectrogram in $p$-th parent-spectrogram is $\frac{M+2L}{2^N} \times \frac{M+2L}{2^N} \times 3$. Note that $N \in \{1, 2, \ldots \log_2^M\}$, $s \in \{1, 2, 3, 4 \}$, $p \in \{1, 2 ,\ldots, 4^{N-1} \}$, and the inverse transformation of the wavelet packet decomposition (IWPD) is denoted as $\mathcal W^{-N}(\x)$.

\subsection{Problem Formulation}
\paragraph{\textbf{Notations}}
\normalsize
We denote a deep learning based image classifier as $f_{\btheta}: \mathcal{X} \rightarrow \mathcal{Y}$, where $\btheta$ is the model parameter, $\X \in \mathbb{R}^3$ and $\Y = \{1, 2, \ldots, C\}$ indicate the input and output space, respectively. The benign training dataset is denoted as $\D = \{(\x^{(i)}, y^{(i)})\}_{i=1}^n$, where $\x^{(i)} \in \X$ denotes a benign image, while $y^{(i)}$ is its ground-truth label. It will be split to two parts, including $\D_c$ and $\D_p$, and the images in $\D_p$ are selected to generate poisoned images, \ie, $(\x^{(i)}, y^{(i)})$ is modified to $(\hat{\x}^{(i)}, t)$ by inserting a trigger \textit{$\triangle$} into $\x^{(i)}$ and changing $y^{(i)}$ to the target label $t$. The set of poisoned image pairs is denoted as $\hat{\D}_p = \{(\hat{\x}^{(i)}, t) | (\x^{(i)}, y^{(i)}) \in \D_p \})$. 
Then, the poisoned training dataset is $\hat{\D} = \{\D_c, \hat{\D}_p\}$, 
and the poisoning ratio is $p = \frac{|\hat{\D}_p|}{|\hat{\D}|}$.
\paragraph{\textbf{Threat Model}}
\vspace{0.2cm}
\normalsize
In this work, we consider the data-poisoning based backdoor attack, where the attacker can only manipulate the benign training dataset $\D$ to obtain the poisoned training dataset $\hat{\D}$, without access to the training process.
In the testing process, given the model $f_{\btheta}$ trained on $\hat{\D}$ by the user, the attacker can generate the backdoor images by inserting the trigger pattern into the benign inference images, and aims to activate the backdoor in the model $f_{\btheta}$, \ie, $f_{\btheta}(\hat{\x}) = t$. 
\vspace{-0.2cm}
\begin{figure}[htbp]
\centering
\subfigure[Image]{
\includegraphics[width=0.1\textwidth]{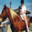}
\label{img}
}
\hspace{0.8cm}
\subfigure[\hspace{-0.05cm}Image without low frequency]{
\label{img_without_low}
\includegraphics[width=0.1\textwidth]{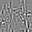}
}
\hspace{0.8cm}
\subfigure[Image without high frequency]{
\label{img_without_high}
\includegraphics[width=0.1\textwidth]{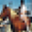}
}
\caption{Impact of high and low frequency information on image vision.}
\label{The influence of high and low frequency information on image vision}
\end{figure}
\vspace{-0.5cm}
\subsection{Analysis on Dataset}
\label{Analysis on dataset}
\paragraph{\textbf{Distribution of Frequency Information in the Dataset}}
In order to make the attack methods more stealthy and precise, we analyze the distribution of frequency information in the dataset to infer the critical frequency regions that DNNs would focus on. These critical frequency regions will then be selected as the poisoning regions for our attack. By employing N-level WPD, we decompose the frequency information of training samples in the dataset, resulting in $4^N$ sub-spectrograms within $4^{N-1}$ parent-spectrograms, where each parent-spectrogram encompasses 4 sub-spectrograms. To quantify the information importance of each sub-spectrogram, we calculate the effectiveness $E$ of the frequency information within each sub-spectrogram across all training samples of dataset $\D$:
% by Eq.~\eqref{effectiveness of the frequency information}: 
% As defined in Eq.~\eqref{effectiveness of the frequency information}, $E$ represents the absolute average magnitude of the absolute average of the coefficient matrix within each sub-spectrogram. 
% This approach helps to mitigate the influence of noise and highlight the regions containing the most ignificant information.
% \begin{equation}
% \label{effectiveness of the frequency information}
%  % E^N_{p,s} = {\left |\text{avg}( \{\sum\limits_{\x^{(i)}\in \D}^{}{\mathcal{W}^{N}_{p,s}(\x^{(i)})\}} ) \right |}
%  \begin{aligned}
% &E^N_{p,s} = {\left |{\frac{1}{m \cdot n \cdot 3} \sum_{i=1}^{m} \sum_{j=1}^{n} \sum_{c=1}^{3}} C^N_{p,s} \right |},  \\ 
% &where \hspace{0.4cm} C^N_{p,s} = \sum\limits_{\x^{(i)}\in \D}^{}{\left[\mathcal{W}^{N}_{p,s}(\x^{(i)})\right]}.
% \end{aligned}
% \end{equation}
\begin{equation}
\label{effectiveness of the frequency information}
 % E^N_{p,s} = {\left |\text{avg}( \{\sum\limits_{\x^{(i)}\in \D}^{}{\mathcal{W}^{N}_{p,s}(\x^{(i)})\}} ) \right |}
 \begin{aligned}
E^N_{\textbf{p},\textbf{s}} = {\left |{\frac{1}{m \cdot n \cdot 3} \sum\limits_{\x^{(i)}\in \D} \sum_{i=1}^{m} \sum_{j=1}^{n} \sum_{c=1}^{3}} C^N_{\textbf{p},\textbf{s}}(\x^{(i)}) \right |},  
\end{aligned}
\end{equation}
\noindent 
In Eq.~\eqref{effectiveness of the frequency information}, $C^N_{\textbf{p},\textbf{s}}$ is a $m \times n \times 3$ coefficient matrix representing the $\textbf{s}$-th sub-spectrogram in $\textbf{p}$-th parent-spectrogram, where the values of $m$ and $n$ given by $\frac{M+2L}{2^N}$.
Considering that the images in the training dataset might be affected by various factors during the acquisition process, such as sensor limitations, transmission errors, or environmental factors, noise may be introduced. Noise usually has a lower amplitude. Consequently, its contribution to the absolute average value is relatively minor compared to that of the actual signal (the related proof will be shown in Supplementary~\ref{explain why it is important to take the average value before taking the absolute value}). This implies that sub-spectrograms with higher $E$ values are more likely to contain significant information or features of interest to the models, rather than being predominantly composed of noise. This understanding is crucial for effectively analyzing and extracting critical features from the dataset.
As a result, we refer to the sub-spectrogram with the largest $E$ value within each parent-spectrogram as the critical frequency region.
% A larger $E$ indicates the dataset has a higher presence of components within the corresponding sub-spectrogram, and we refer such sub-spectrogram as the critical frequency regions. 
It should be noticed that when conducting statistical analyses, we do not take into account the information of the lowest frequency. Fig.~\ref{The influence of high and low frequency information on image vision} shows that the lowest frequency information always plays an impact on image in visual, so we refrain from making any alterations to it.

% We remove the low frequency and high frequency information of an image respectively, and Fig.~\ref{The influence of high and low frequency information on image vision} shows that the lowest frequency information always 
% % contains the most information and
% plays an impact on image in visual. It is difficult to maintain the visual naturalness after modifying the lowest frequency. Therefore, we do not take into account the information of the lowest frequency when we conduct statistical analysis. 
% \begin{remark}
% $E^N_{p,s}$ is represented as ${\left | \text{avg}(C^N_{p,s})  \right |}$ instead of ${ \text{avg}(\left |C^N_{p,s}\right |)}$. Specially, the elements in $C^N_{p,s}$ include positive and negative values. 
% %Specially, the elements in $C^N_{p,s}$ include positive and negative values. Positive values indicate that the image signal energy is gathered in the direction consistent with the wavelet basis function, and negative values indicate that the image signal energy is gathered in the reverse direction. Considering the direction, we average the coefficient matrix before taking the absolute.
% \end{remark}
% \vspace{-0.5cm}
\begin{figure}[htp]
\centering
\includegraphics[width=8cm,height =2.2cm]{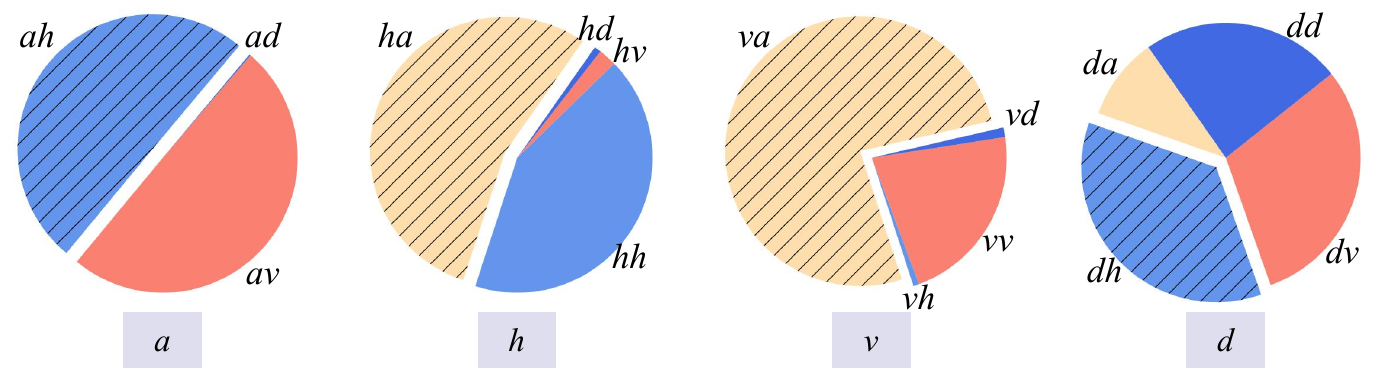}
% figures/z_bar.pdf
\caption{Frequency information distribution of CIFAR-10 in \textit{`a'}, \textit{`h'}, \textit{`v'}, \textit{`d'} parent-spectrogram, respectively.}
\label{CIFAR-10-distribution}
\end{figure}
 
\begin{algorithm}[H]
\caption{Selection of Critical Frequency Regions to Mask}\label{The algorithm of WPDA}
\begin{algorithmic}[1]
\setlength{\baselineskip}{0.95 \baselineskip}
\REQUIRE The benign training dataset as $\D = \{\x^{(1)}, \dots, \x^{(n)}\}$, the shape of each sample is $M \times M \times 3$, the level of WPD is $N$, and the vanishing moment of the wavelet basis function is $2L$.
% $\mathcal{W}$ denotes the WPD .
\ENSURE The list of critical frequency regions $k_{list}$ that will be masked in the training samples.
 % of s-th sub-spectrogram in p-th parent-spectrogram.

% \FOR{$p\leftarrow 1$ \TO $4^{N-1}$} 
\STATE{$k_{list}=[]$}
\FOR{$\textbf{p} = {1, \dots, 4^{N-1}}$} 
    \STATE{$\textbf{s}_{list}=[]$}
    \FOR{$\textbf{s} = 1$ \TO 4} 
        % \FOR{$s$ in range 4} \do
        \STATE $E^N_{\textbf{p},\textbf{s}} = {\left |{\frac{1}{m \cdot n \cdot 3} \sum\limits_{\x^{(i)}\in \D} \sum_{i=1}^{m} \sum_{j=1}^{n} \sum_{c=1}^{3}} C^N_{\textbf{p},\textbf{s}}(\x^{(i)}) \right |}$
        % {$C^N_{p,s} = \sum\limits_{\x^{(i)}\in \D}^{}{\left[\mathcal{W}^{N}_{p,s}(\x^{(i)})\right]}$}
        % {${C^N_{p,s}} = \{\sum\limits_{x^{(i)}\in \D}^{}{\mathcal{W}^{N}(x^{(i)})\}_{p,s}}$}
        % \STATE{$E^N_{p,s} = {\left |{\frac{1}{m \cdot n \cdot 3} \sum_{i=1}^{m} \sum_{j=1}^{n} \sum_{c=1}^{3}} C^N_{p,s} \right |}$, \\$where \hspace{0.3cm} m = n = \frac{M+2L}{2^N}$}
        % {$E^N_{p,s} = \left |\text{avg}(C^N_{p,s})  \right |$}
        \STATE {$\textbf{s}_{list} \leftarrow \textbf{s}_{list} \cup \{(E^N_{\textbf{p},\textbf{s}},\textbf{p},\textbf{s})\}$}
        \ENDFOR
    \IF{$\textbf{p}=1$}{
        \STATE{$ \textbf{s}_{list} \leftarrow \textbf{s}_{list} \setminus \mathop{\arg\max}\limits_{(E^N_{\textbf{p},\textbf{s}}, \textbf{s}, \textbf{p}) \in \textbf{s}_{\text{list}}} E^N_{\textbf{p},\textbf{s}}$} 
        % {$s_{list} \leftarrow s_{list} \setminus \{\max(s_{{list}})\}$}
        % {$s_{list}={s_{list}}\text{.remove}(\max(s_{list}))$} 
        \STATE{$\textbf{p}, \textbf{s}^*=\mathop{\arg\max}\limits_{(E^N_{\textbf{p},\textbf{s}}, \textbf{s}, \textbf{p}) \in \textbf{s}_{\text{list}}}E^N_{\textbf{p},\textbf{s}}$}}
    \ELSE{
        \STATE{$\textbf{p}, \textbf{s}^*=\mathop{\arg\max}\limits_{(E^N_{\textbf{p},\textbf{s}}, \textbf{s}, \textbf{p}) \in \textbf{s}_{\text{list}}}E^N_{\textbf{p},\textbf{s}}$}}
    \ENDIF
    \STATE {$k_{list} \leftarrow k_{list}\cup \{(\textbf{p}, \textbf{s}^*)\}$}

   \ENDFOR
   \RETURN $ k_{list}$
\end{algorithmic}
\label{alg1}
\end{algorithm}
 
%%%%%%%%%%%%%%%%%%%%%%%%%%%%%

\paragraph{\textbf{Differences Among Datasets}}
The complexity of the dataset is affected by the number of categories, image volume, \etc~ The datasets with more categories or larger image volume tend to be more complex and are required to be analysed with more in-depth decomposition, \textit{i.e.}, a higher level WPD.
We regard the sub-spectrograms in each parent-spectrogram with most effective information as the critical frequency regions. Take CIFAR-10 as an example, Fig.~\ref{CIFAR-10-distribution} illustrates the frequency information distribution of CIFAR-10 decomposed by 2-level WPD, and the parent-spectrograms are the frequency regions in the 1-level WPD, \textit{i.e.}, \textit{`a'}, \textit{`h'}, \textit{`v'}, \textit{`d'}. The lowest frequency \textit{`aa'} is not counted in the statistics. As a result, the sub-spectrograms of \textit{`ah'}, \textit{`ha'}, \textit{`va'}, \textit{`dh'} in each parent-spectrograms, containing the most effective information, will be selected as poisoning frequency regions. It is worth noting that the asymmetry between vertical and horizontal components in the dataset is a common phenomenon, resulting from the physical properties of images.
Furthermore, complex datasets require more in-depth decomposition. The details are shown in Algorithm~\ref{alg1}.

% \begin{figure*}[!ht]
% \centering
% \includegraphics[width=17cm,height =6.3cm]{figures/overview10.pdf}
% \caption{Overview of WPDA. 
% \textit{Training Dataset Analysis} module describes the process of analyzing the frequency information distribution of the training dataset using the absolute average algorithm. This analysis identifies the most effective sub-spectrogram within each parent-spectrogram, which are then designated as the poisoning frequency regions. \textit{Backdoor Dataset Generation} module describes the generation process for poisoned samples, which is based on wavelet packet decomposition.
% }
% \label{overview}
% \end{figure*}
\begin{figure*}[!ht]
\centering
\includegraphics
% [width=15cm,height =7.2cm]
[width=17cm,height =7.87cm]
{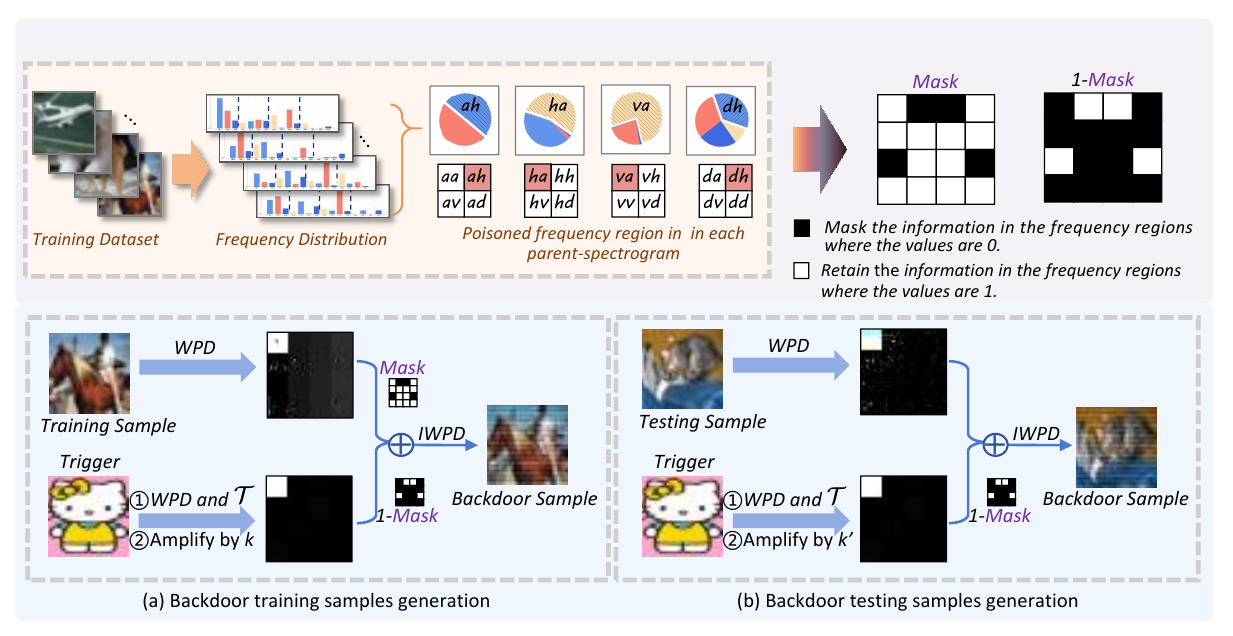}
\caption{Process of WPDA on poisoned samples generation. Before inserting the trigger, we perform frequency distribution statistics on all training samples based on WPD to identify the most critical sub-spectrogram in each parent-spectrogram as the poisoning regions (disregarding the lowest frequency region), and then generate the mask for the poisoned samples generation. In the process of poisoned samples generation, $\mathcal{T}$ represents average transformation.
% \caption{Generation of poisoned training samples for WPDA.
}
\label{overview}
\end{figure*}

\subsection{Proposed Method}
\label{method}
Here we elaborate the proposed WPD-based backdoor attack, dubbed \textbf{WPDA}. Since we focus on data-poisoning based attack, we only describe how to generate poisoned samples shown in Fig.~\ref{overview}:
\begin{enumerate}[]
\item  \textbf{Generating the Frequency-Based Trigger}. 
A trigger image with the shape $M \times M \times 3$ is converted from the spatial domain to the frequency domain via WPD, resulting in $4^N$ sub-spectrograms. Furthermore, we apply an average transformation $\mathcal T$ to replace the elements in each sub-spectrogram by the average of the coefficient matrix in the corresponding sub-spectrogram. The frequency-based trigger $\triangle_{fre}$ is denoted as follows:
% \begin{equation}
% \label{trigger_in_frequency}
%  % \mathcal T(W^N(\triangle))= \bigcup_{s=1}^{4}\bigcup_{p=1}^{4^{N-1}}\left \{\text{avg}(\left \{\mathcal W^N(\triangle)\right \}_{p,s}) \cdot \mathbf{J}\right \}_{p,s},
% \triangle_{fre} = \sum_{i=1}^{4^{N}} \left[{( {{\frac{1}{m \cdot n \cdot 3} \sum_{i=1}^{m} \sum_{j=1}^{n} \sum_{c=1}^{3}} \left[\mathcal{W}^{N}_{p,s}(\triangle) \right ]}) } \cdot \mathbf{J}\right ].
% \end{equation}
% \begin{equation}
% \label{trigger_in_frequency2}
% \overline{C^N_{p,s}}(\triangle)=\mathcal T(C^N_{p,s}(\triangle))
%  =  \mathbf{J} \cdot {( {{\frac{1}{m \cdot n \cdot 3} \sum_{i=1}^{m} \sum_{j=1}^{n} \sum_{c=1}^{3}} C^N_{p,s}(\triangle) }) }.
% \end{equation}
\begin{equation}
\label{trigger_in_frequency}
\begin{aligned}
&\overline{C^N_{\textbf{p},\textbf{s}}}(\triangle)
=\mathcal T(C^N_{\textbf{p},\textbf{s}}(\triangle))
 =  \mathbf{J} \cdot {( {{\frac{1}{m \cdot n \cdot 3} \sum_{i=1}^{m} \sum_{j=1}^{n} \sum_{c=1}^{3}} C^N_{\textbf{p},\textbf{s}}(\triangle) }) },\\
&\triangle_{fre}
= \begin{bmatrix}
  \overline{C^N_{1,1}}(\triangle))  & \overline{C^N_{1,2}}(\triangle)  & \dots  & \overline{C^N_{2^{N-1},2}}(\triangle) \\
  \overline{C^N_{1,3}}(\triangle)  & \overline{C^N_{1,4}}(\triangle)  & \dots  & \overline{C^N_{2^{N-1},4}}(\triangle) \\
  \vdots & \vdots & \overline{C^N_{\textbf{p},\textbf{s}}}(\triangle) & \vdots \\
  \vdots  & \vdots  & \dots  & \overline{C^N_{4^{N-1},4}}(\triangle)
\end{bmatrix}.
% ,\\
 % \mathrm{s.t.}\quad & 0\le i=j \le {M+2L-1},~0\le m=n \le M-1.
\end{aligned}
\end{equation}
In Eq.~\eqref{trigger_in_frequency}, $\triangle$ represents the trigger image, $\mathbf{J}$ is a matrix of ones with the same shape as the coefficient matrix in the sub-spectrogram, \textit{i.e.}, $\frac{M+2L}{2^N} \times \frac{M+2L}{2^N} \times 3$, and the values of $m$ and $n$ are calculated from $\frac{M+2L}{2^N}$. Average transformation $\mathcal T$ preserves the original information of the trigger image while improving its visual stealthiness. Fig.~\ref{stealthy} presents the results of the differences between the original samples and the poisoned samples in the CIFAR-10 dataset, under conditions with and without the average transformation $\mathcal T$. Based on mean square error (MSE) and structural similarity index measure (SSIM), we demonstrate that average transformation $\mathcal T$ can effectively improve the stealthiness of the trigger in visual. 

\begin{figure}[htbp]
\centering
\includegraphics[width=8.0cm,height =3.0cm]
{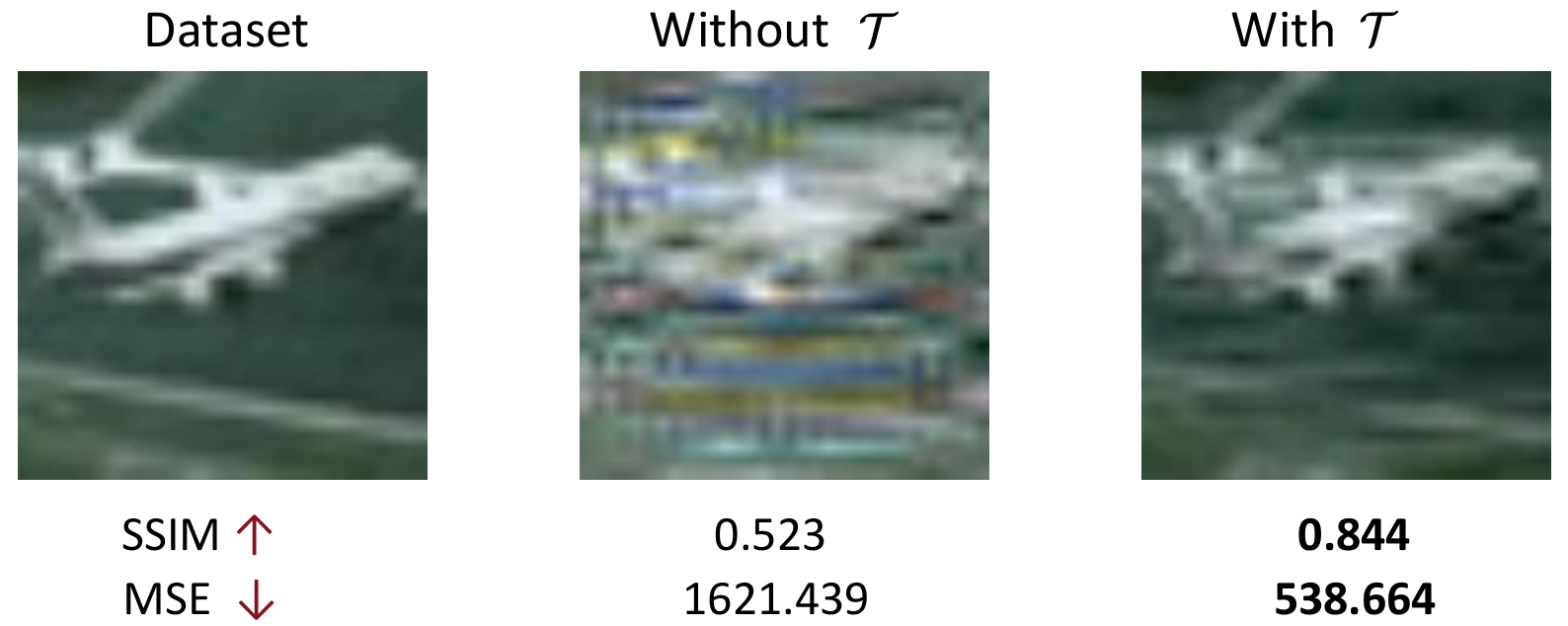}
\caption{The benign samples and the poisoned samples generated without and with average transformation $\mathcal T$. SSIM and MSE values are the average measurements calculated from training samples. Higher SSIM values and lower MSE values indicate greater similarity between the poisoned samples and original samples.}
\label{stealthy}
\end{figure}

\item  \textbf{Generating the Poisoned Training Samples}. Following Algorithm~\ref{alg1}, we select $4^{N-1}$ critical frequency regions identified as the most critical portions of the parent-spectrograms, and these regions are then designated as poisoning frequency regions for the attack. The generation of poisoned training samples, denoted as $\hat{\boldsymbol{x}}_{train}$, \textit{i.e.}, $g({x}_{train};k)$, is then performed according to Eq.~\eqref{poisoned_training_samples_generation_equation}: 

\begin{equation}
% \footnotesize
\begin{aligned}
\label{poisoned_training_samples_generation_equation}
g({\x}_{train};k) = \mathcal W^{-N}(\mathcal W&
^N (\x_{train})\odot \mathbf{m} + k \cdot \triangle_{fre} &  \\
\odot (\mathbf{1-m})),
\end{aligned}
\end{equation}
where $k$ is the trigger intensity of the poisoned training samples $\hat{\boldsymbol{x}}_{train}$, $\mathbf{m} \in \left \{ 0,1 \right \} ^{(M+2L) \times (M+2L) \times 3}$ is a binary mask, the elements of $\mathbf{m}$ in poisoning frequency regions are 0, and $\odot$ represents element-wise multiplication of matrices.

\item \textbf{Generating the Poisoned Testing Samples}.
In practical application scenarios, we need to consider the visual performance of the poisoned testing samples, so we retain the information of the original image with the intensity of $\alpha$ in the poisoning frequency regions in the process of generating the poisoned testing samples $\hat{\boldsymbol{x}}_{test}$, \textit{i.e.}, $g'({x}_{test};k')$.
The poisoned testing samples are generated as Eq.~\eqref{poisoned_test_samples_generation_equation}:
\begin{equation}
% \footnotesize
\label{poisoned_test_samples_generation_equation}
\begin{aligned}
{g'}({\x}_{test};{k'}
) 
&=\mathcal W^{-N} (\mathcal {W}^N (\x_{test})\odot \mathbf{m} + \mathbf{(1-m)} \odot\\& 
(\alpha  \cdot \mathcal {W}^N (\x_{test}) + {k'}  \cdot \triangle_{fre})) ,
\end{aligned}
\end{equation}
where 
% $\hat{\boldsymbol{x}}_{test} = g'({x}_{test};k')$, 
$k'$ is the trigger intensity and $\alpha$ is the original benign sample information intensity of the poisoned testing samples $\hat{\boldsymbol{\x}}_{test}$.
\end{enumerate}
\vspace{-0.3cm}

\begin{table}[ht]
\caption{Notation definitions.}
\renewcommand\arraystretch{1.1}
\centering
\setlength{\tabcolsep}{2pt}
\scalebox{0.8}{
\begin{tabular}{c|l}
\toprule
Notation  &  Definition  
\\ 
\hline
$\x$  & One data point, vector  
\\ 
$y$  &  One label, scalar  
\\ 
$f_{\btheta}$  &  Model with parameter
\\ 
$\btheta$  &  Model parameters  
\\ 
$f_{\btheta}(\x)$ & Model output
\\
$\mathcal{X} \in \mathbb{R}^3$  & $3$-dimensional input space 
\\
$\mathcal{Y} \in \{1, 2, \ldots, C\}$ & Output space with $C$ candidate classes
\\
$\D = \{(\x^{(i)}, y^{(i)})\}_{i=1}^n$  &  Dataset with $n$ samples 
\\ 
$\hat{\D}_p = \{(\hat{\x}^{(i)}, t) | (\x^{(i)}, y^{(i)}) \in \D_p \})$  & Set of poisoned sample pairs  
\\ 
$\hat{\D} = \{\D_c, \hat{\D}_p\}$  & Poisoned
training dataset 
\\
$p = \frac{|\hat{\D}_p|}{|\hat{\D}|}$  & Poisoning ratio 
\\
$\triangle$  & Trigger image in spatial domain 
\\
$\triangle_{fre}$  & Trigger image in frequency domain 
\\
$N$  & Level of WPD 
\\
$\mathcal W^{N}(\cdot)$  & $N$-level WPD
\\
$C^N_{\textbf{s},\textbf{p}}=\mathcal W^{N}_{\textbf{s},\textbf{p}}(\cdot)$  & \multicolumn{1}{l}{A coefficient matrix representing the $\textbf{s}$-th sub-spectrogram }\\& \multicolumn{1}{l}{in the $\textbf{p}$-th parent-spectrogram}
\\
$\mathcal W^{-N}(\cdot)$  & $N$-level IWPD
\\
$\mathcal{T}$  & Average transformation
training dataset 
\\
$\mathbf{m} \in \left \{ 0,1 \right \} ^{(M+2L) \times (M+2L) \times 3}$  & A binary mask with the shape of ${(M+2L) \times (M+2L) \times 3}$
\\
$2L$  & The vanishing moment of the wavelet basis function
\\
$\odot$  & Element-wise multiplication of matrices
\\
$k$  & \multicolumn{1}{l}{The intensity of the trigger image information $\triangle_{fre}$ in} \\& \multicolumn{1}{l}{the poisoned training samples $\hat{\boldsymbol{x}}_{train}$}
\\
$k'$  & \multicolumn{1}{l}{The intensity of the trigger image information $\triangle_{fre}$ in} \\& \multicolumn{1}{l}{the poisoned testing samples $\hat{\boldsymbol{x}}_{test}$}
\\
$\alpha$  & \multicolumn{1}{l}{The intensity of the original benign sample information in} \\& \multicolumn{1}{l}{the poisoned testing samples $\hat{\boldsymbol{x}}_{test}$}
\\
\bottomrule
\end{tabular}%
}
\label{table: template}
\vspace{-0.5cm}
\end{table}

\subsection{Characteristic Analyses}
\label{Qualitative Analysis}
% \paragraph{Characteristic Definitions}
A great-performance backdoor attack should simultaneously possess stealthiness, effectiveness, and resistance. These three characteristic definitions are as follows:
\begin{itemize}
\item  \textbf{Stealthiness}: The stealthiness of backdoor attacks mainly manifests in two aspects: \ding{172}The trigger stealthiness during training process could be achieved with a low poisoning ratio (\textit{i.e.}, a small value of $p$). A small number of poisoned training samples mixed into the dataset not only makes human inspection costly but also difficult for existing poisoned sample detection methods to identify.
\ding{173}The trigger stealthiness during testing process could by achieved by ensuring visual naturalness of the poisoned testing samples, \textit{i.e.}, the poisoned testing samples look similar to the original testing samples. To achieve this goal, in addition to adopting average transformation $\mathcal{T}$ (see Fig.~\ref{stealthy} and Eq.~\ref{trigger_in_frequency}), it can also be further improved by controlling the hyperparameters of the original benign sample information $\alpha$ and trigger intensity $k'$ in the poisoned testing samples. In Eq.~\ref{poisoned_test_samples_generation_equation}, the original benign sample information intensity $\alpha$ is desired to be set a large value and the trigger intensity $k'$ in the poisoned testing samples is desired to be set a small value, making the trigger invisible in the poisoned testing samples.

% \textbf{Stealthiness}: The stealthiness of backdoor attacks mainly manifests in two aspects: \ding{172} Backdoor attacks achieve success with a low poisoning ratio. High poisoning ratios exacerbate the differences between poisoned and benign samples, which backdoor detection techniques like AC and SCAn exploit to identify poisoned samples. Therefore, a small amount of data poisoning is inherently more stealthy, making it difficult for backdoor detection techniques to accurately filter out poisoned training samples.~\cite{wu2022backdoorbench}
% \ding{173} The poisoned testing samples are natural in visual. We improve the visual naturalness of the poisoned testing samples by adopting average transformation $\mathcal{T}$, retaining the original benign sample information in the poisoned frequency region, \textit{i.e.}, $\alpha = 1$, and limiting trigger intensity $k'$ in the poisoned testing samples to prevent it from being too high.
\item  \textbf{Effectiveness}: The effectiveness of backdoor attacks mainly manifests in two aspects: \ding{172} The trigger could be effectively injected into the model through training process. To achieve this goal, a high poisoning ratio $p$ and a high trigger intensity $k$ in the poisoned training samples are desired. Since our study focus on the case of low poisoning ratio, it can be improved by controlling $k$.
\ding{173} The trigger could be effectively activated during testing process. According to Eq.~\ref{poisoned_test_samples_generation_equation}, a small value of original benign sample information $\alpha$ and a large value of trigger intensity $k'$ in the poisoned testing samples are desired.
% \item  \textbf{Effectiveness}: The effectiveness of backdoor attacks mainly manifests in two aspects: \ding{172} The model effectively learns trigger information rather than memorising the poisoned training samples. During the process of backdoor injection, a high poisoning ratios and high trigger intensity $k$ in the poisoned training samples conducive to the model effectively learn the trigger information. However, under the constraint of a low poisoning ratio, a higher trigger intensity $k$ in the poisoned training samples is required for the model to effectively learn the trigger information.
% \ding{173} Triggers in the poisoned testing samples can effectively activate the backdoor. During the process of trigger activation, a low intensity of original benign sample information $\alpha$ and high trigger intensity $k'$ in the poisoned testing samples conducive to the model precisely activate the backdoor. 
\item  \textbf{Resistance}:
Resistance means that a successful backdoor attack could bypass the backdoor defenses. According to ~\cite{wu2022backdoorbench}, there are two kind of backdoor defenses, \textit{i.e.}, in-training defenses and post-training defenses.
\ding{172}In-training defenses aim to explore the differences between benign samples and poisoned samples during the training process. To the evaluation of~\cite{wu2022backdoorbench}, a high poisoning ratio $p$ and a high trigger intensity $k$ in the poisoned training samples could highlight the differences between benign samples and poisoned samples, which means that strong poisoning accelerates the learning speed of the poisoned samples. 
\ding{173}Post-training defenses aim to remove or modify the values of the backdoor-related neurons. To the evaluation of~\cite{wu2022backdoorbench}, a high poisoning ratio $p$ and a high trigger intensity $k$ in the poisoned training samples could enhance the backdoor effect, indicating that the backdoor neurons are easy to be identified. In short, a low poisoning ratio $p$ and a low trigger intensity $k$ in the poisoned training samples are desired.
% \item  \textbf{Resistance}: The resistance of backdoor attacks mainly manifests in two aspects: Trigger information in the poisoned training samples is required not to be too prominent. \ding{172} There is a loss gap between the poisoned and benign samples, which some training-based defense techniques exploit to identify poisoned samples. The higher the trigger intensity $k$ in the poisoned training samples, the larger the gap.
 % \ding{173} There are differences in the neurons activated between the poisoned and benign samples, which some post-training-based defense techniques exploit to identify poisoned samples and poisoned neurons. The higher the trigger intensity $k$ in the poisoned training samples, the more accurately the poisoned neurons are pruned. Therefore, a low trigger intensity $k$ in the poisoned training samples is required for resistance.~\cite{wu2022backdoorbench}
\end{itemize}
% \begin{figure}[htbp]
% \vspace{-0.3cm}
% \centering
% \includegraphics[width=5.3cm,height =3.1cm]{figures/zcifar10_intensity_111.pdf}
% \caption{Impact of the trigger intensity $k$ of the poisoned training samples on model performance.
% % on  PreAct-ResNet18 and VGG19-BN trained by CIFAR-10 with different poisoning ratios, and use poisoned vaildation set to test the accuracy of trigger activation.
% }
% \label{prominence}
% \end{figure}
% In Summary, according to above analyses, we found that stealthiness requires a low $k'$ and a high $\alpha$, effectiveness requires a high $k$ and a high $k'$, and resistance requires a low $k$. 

Obviously, there are conflicts among characteristics. To resolve these conflicts, we meticulously design the following strategy:
\begin{itemize}
\item {\textbf{Stealthiness \& Effectiveness}}: According to above analyses, we found that stealthiness requires a high $\alpha$ and a low $k'$, while effectiveness requires a low $\alpha$ and a high $k'$. To ensure stealthiness, we retain the original benign sample information in the poisoned testing samples, \textit{i.e.}, $\alpha$ is set to 1. We focus on the effect of the value of $k$ and $k'$. According to~\cite{wu2022backdoorbench}, the trigger has generalization capability. We evaluate the trigger generalization capability with CIFAR-10, and generate poisoned testing samples from validation set. Fig.~\ref{test_generation3} illustrates trigger activation accuracy of the backdoored models with different $k$ and $k'$ values at a 0.004\% poisoning ratio. A low $k$ usually requires a high $k'$ to successfully activate the trigger. Fig.~\ref{test_generation2} illustrates the trigger activation accuracy with different $k'$ at a fixed $k$ across different poisoning ratios. At a low poisoning ratio, a high $k'$ facilitate the effective trigger activation. \textbf{Specifically, in our setting, considering both stealthiness and effectiveness, we require $k$ to be a moderate value, and $k'$ to be slightly smaller than $k$ or slightly larger than $k$ to ensure effective trigger activation.} (Notation definitions are shown in Tab.~\ref{table: template}.)
% \textit{i.e.}, $k'$ value should be set a slightly smaller value or a larger value than $k$ to ensure effectiveness. We evaluate the trigger generalization capability with CIFAR-10, and generate poisoned testing samples from validation set. Fig.~\ref{k' generation} illustrates the effect of the trigger intensity $k'$ in the poisoned testing samples on trigger activation accuracy. A high $k'$ facilitate the the trigger activation.  Specially, in our setting, considering stealthiness and effectiveness, we require $k$ to be a moderate value, and $k'$ to be slightly smaller than $k$ or slightly larger than $k$.

\begin{figure}[htbp]
\centering
\subfigure[Effect of $k$ and $k'$ values on trigger activation accuracy at a 0.004\% poisoning ratio.
]{
\includegraphics[width=1.50in]
{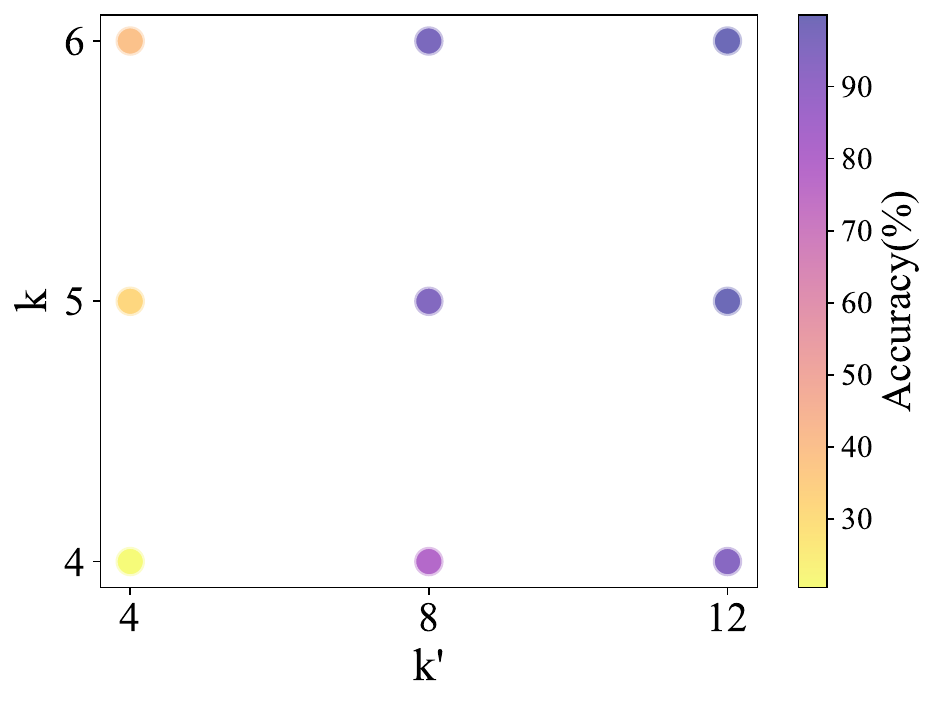}
\label{test_generation3}
}
\hspace{3mm}
\subfigure[Effect of $k'$ values at different poisoning ratios on trigger activation accuracy with a fixed $k$.
]{
\includegraphics[width=1.50in]
{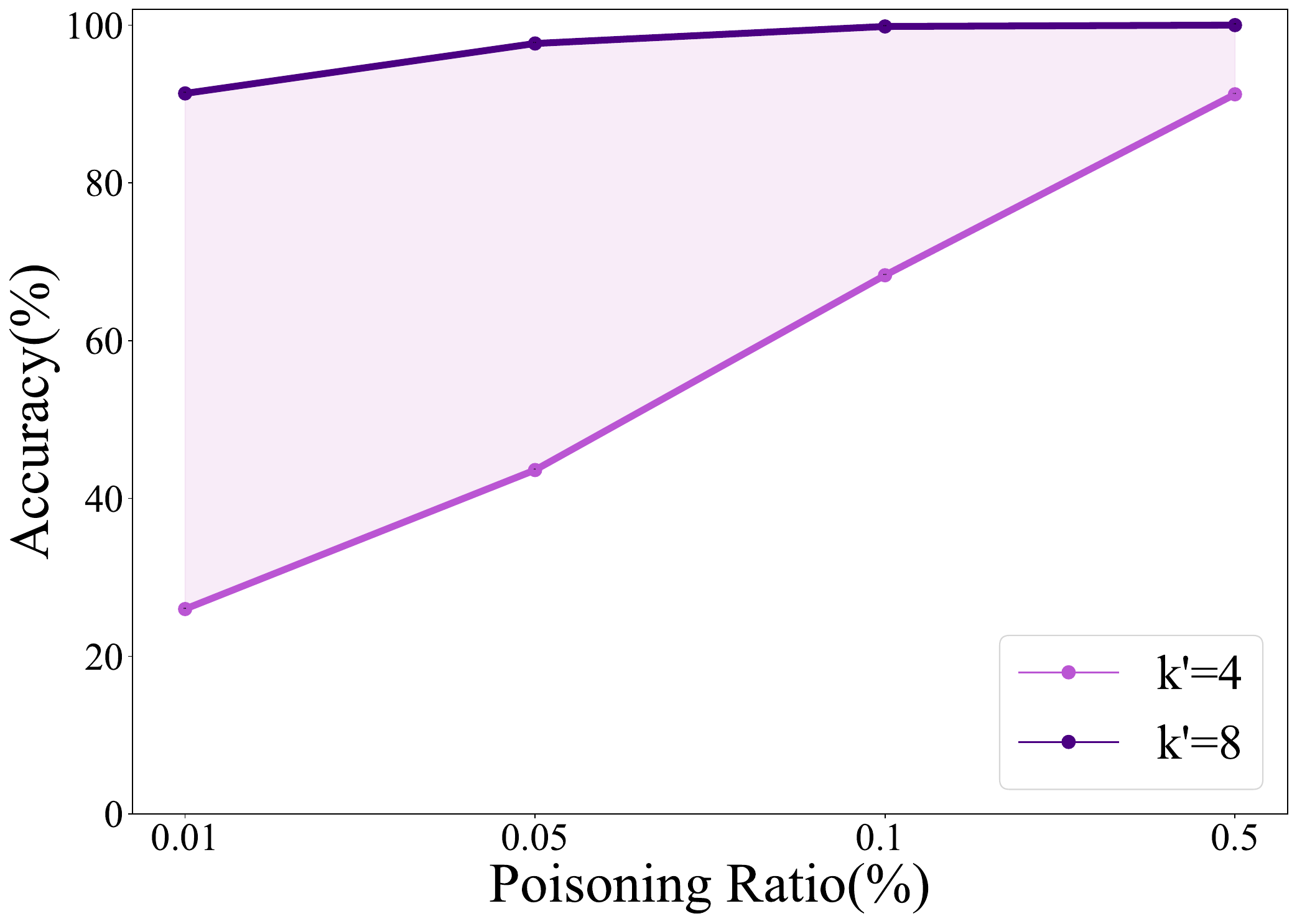}
\label{test_generation2}
}
\caption{Effect of the trigger intensity on trigger activation accuracy.
}
\label{fig:generation}
\end{figure}
% \begin{figure}[!ht]
% \centering  %图片全局居中
%     {
%     \includegraphics[width=1.60in]{figures/test_generation3.pdf}}
%     \label{test_generation1}
%     {
%     \includegraphics[width=1.60in]{figures/test_generation2.pdf}}
%     \label{test_generation2}
% \caption{Effect of the trigger intensity on trigger activation accuracy.}
% \label{k' generation}
% \end{figure}
\item {\textbf{Effectiveness \& Resistance}}: According to above analyses, we found that effectiveness requires a high $k$ and a high poisoning ratio $p$, while resistance requires a low $k$ and a low poisoning ratio $p$. \textbf{Since our study focus on the case of low poisoning ratio, although a small $k$ is desired,  at the meantime, it must also ensure effective trigger injection during the training process.}  (Notation definitions are shown in Tab.~\ref{table: template}.)

\end{itemize}

\section{Experiments}
\label{Experiment}
\subsection{Experiment Settings}
\begin{figure}[htbp]
\vspace{-0.3cm}
\centering
\includegraphics[width=5.3cm,height =3.1cm]{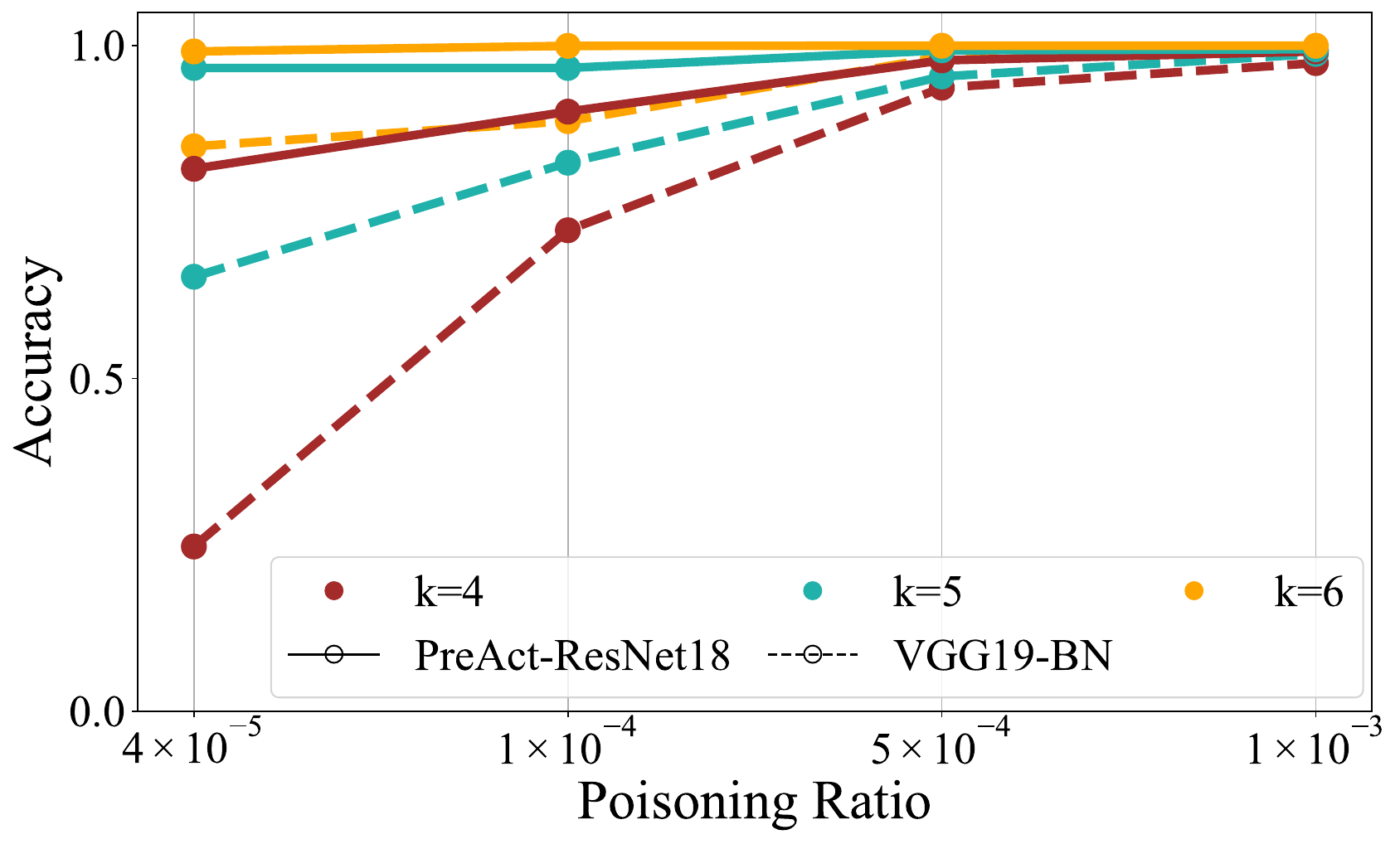}
\caption{
% Impact of the trigger intensity $k$ of the poisoned training samples on model performance. 
Evaluation of backdoored models effectiveness in learning trigger at different trigger intensities.  
}
\label{prominence}
\end{figure}

\paragraph{\textbf{Datasets and Models}}
% \noindent\textbf{Datasets and Models.} 
We conduct experiments on three benchmark datasets, including CIFAR-10~\cite{2009Learning}, CIFAR-100~\cite{2009Learning} and Tiny ImageNet~\cite{le2015tiny}. Specifically, CIFAR-10~\cite{2009Learning} contains 10 classes with 60,000 color images and the size of each image is ${32\times 32 \times 3}$. There are 50,000 images for training and 10,000 images for testing. CIFAR-100~\cite{2009Learning} has 100 classes, each of them containing 500 training images and 100 testing images. The size of an image in CIFAR-100 is the same as that in CIFAR-10~\cite{2009Learning}. Tiny ImageNet~\cite{le2015tiny} is a subset of ImageNet~\cite{deng2009imagenet} which contains 200 classes, with 500 images per class for training, 50 images per class for validation, and 50 images per class for testing. The size of each image is ${64\times 64 \times 3}$. The model architectures that we adopted are PreAct-ResNet18~\cite{he2016identity} and VGG19-BN~\cite{simonyan2014very}. 
\paragraph{\textbf{Attack Settings}}
We compare the proposed method with seven current state-of-the-art (SOTA) backdoor attacks, which can be divided into four categories: 1) \textit{non clean-label with spatial-static trigger}, such as BadNets~\cite{gu2019badnets} and Blended~\cite{chen2017targeted}; 2) \textit{non clean-label with spatial-dynamic trigger}, like SSBA~\cite{li2021invisible} and Input-aware~\cite{nguyen2020input}; 3) \textit{clean-label with spatial-static trigger}, like SIG~\cite{barni2019new}; 4) \textit{non clean-label with frequency-static trigger}, including LF~\cite{zeng2021rethinking} and WABA~\cite{drager2022backdoor}. To verify the effectiveness of our method, we compare the attack performance with the SOTA attacks under different poisoning ratios, ranging from $0.004\%$ to $1\%$. 
% It is worth noting that $0.004\%$ means that there are only 2 poisoned samples in the training dataset of CIFAR-10~\cite{2009Learning} and CIFAR-100~\cite{2009Learning}, and 4 poisoned samples in the training dataset of Tiny ImageNet~\cite{le2015tiny}. 
\paragraph{\textbf{Defense Settings}}
We validate the performance of our method against multiple popular backdoor defense and detection methods. There are three backdoor detection methods (\textit{e.g.}, AC~\cite{chen2018detecting}, SCAn~\cite{tang2021demon}, and STRIP~\cite{gao2019strip}.) and five SOTA backdoor defense methods (\textit{e.g.}, CLP~\cite{zheng2022data}, FT, I-BAU~\cite{zeng2021adversarial}, NAD~\cite{li2021neural}, and NC~\cite{wang2019neural}.) are considered. For the detection methods, SCAn~\cite{tang2021demon} and STRIP~\cite{gao2019strip}, which assume that a detector can access a small set of benign samples, we assume the number of benign samples for each class to be 10, as recommended in these works. For the defense methods, we follow the default configuration used in BackdoorBench~\cite{wu2022backdoorbench,wu2023defenses} for a fair comparison. 
\begin{figure}[htbp]
\centering
\vspace{-0.3cm}
\includegraphics[width=3.4in]{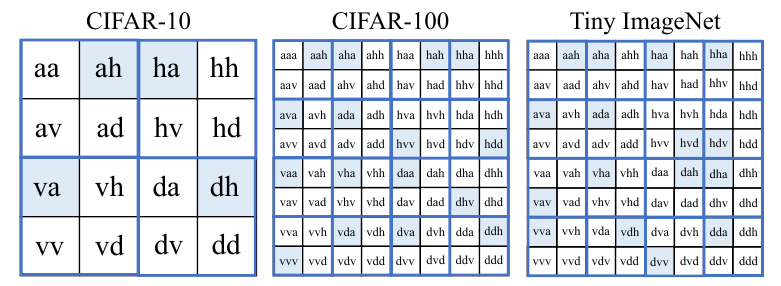}
\caption{The selected regions for CIFAR-10, CIFAR-100, and Tiny ImageNet, respectively.}
\label{selected region}
\end{figure}
\paragraph{\textbf{Details of WPDA}}
According to the description of WPDA shown in Sec.~\ref{Method}, we will provide more details about the decomposition level of dataset, the selection of the poisoning regions, and the design of the trigger. 
\begin{enumerate}
    \item \textbf{Dataset Decomposition}.
Datasets with large volumes and numerous categories are often complex, requiring detailed decompositions to achieve effectiveness at extremely low poisoning ratios. Hence, in the experiment, we perform a 2-level WPD on CIFAR-10, while CIFAR-100 and Tiny ImageNet undergo a 3-level WPD. The wavelet basis function is \textit{`db3'}, and the vanishing moment is $2L=4$.
\item \textbf{Poisoning Regions Selection}.
Based on the analysis in Sec.~\ref{Analysis on dataset}, $4^{N-1}$ key poisoning regions of sub-spectrograms are selected as the poisoning regions (except the lowest frequency region in the spectrogram). We highlight the selected regions in Fig.~\ref{selected region}, in which it can be seen that there are 4 regions chosen in CIFAR-10 and 16 regions in CIFAR-100 and Tiny ImageNet. 
% \textbf{3)} Trigger Design. 
% \textit{3) Trigger Design:} 
\item \textbf{Trigger Design}.
In the following experiment, we select ``Hello Kitty'' image derived from Blended as the trigger. Take CIFAR-10 as an example, we generate poisoned testing samples from vaildation set, and set $k=k'$ to evaluate effectiveness of the backdoored models for learning trigger. Fig.~\ref{prominence} illustrates at a low poisoning ratio 0.004\%, a large $k$ is desired to ensure successful attack. Therefore, based on preliminary experiments with the validation set, $k$ should be at least 6 in CIFAR-10. 
Considering the difference among the datasets, the trigger information intensity $k$ in the poisoned training samples is altering, \eg, $k=6$ in CIFAR-10, $k=8$ in CIFAR-100, and $k=7$ in Tiny ImageNet. The values of the trigger information intensity $k'$ in the poisoned testing samples for different datasets can be referred to Tab.~\ref{test-value}, and the original benign sample information intensity $\alpha$ in the poisoning regions is 1.
\end{enumerate}
% \vspace{-0.1cm}
\begin{table}[!ht]
\centering
\caption{The value of $k'$ in Eq. \eqref{poisoned_test_samples_generation_equation} under the condition of different poisoning ratios on datasets.}
\label{test-value}
\scalebox{0.98}{
\begin{tabular}{lllcccc}
\hline 
\begin{tabular}[c]{@{}l@{}} {Poisoning ratio $\rightarrow$}\\{Dataset $\downarrow$}   \end{tabular}             & \multicolumn{1}{c}{0.004\%} & \multicolumn{1}{c}{0.01\%} & 0.05\% & 0.1\% & 1\% \\ \hline \hline
CIFAR-10                    & 8                           & 8                          & 6                          & 6                          & 6   \\
CIFAR-100                   & 10                           & 8                           & 7                           & 6                           & 5   \\ 
Tiny ImageNet                         & 8                           & 8                           & 7                           & 7                           & 7   \\ \hline
\end{tabular}}
\vspace{-0.2cm}
\end{table}

% \begin{table}[!ht]
% \centering
% \caption{The value of $k'$ in Eq.~\eqref{poisoned_test_samples_generation_equation} under the condition of different poisoning ratios on datasets.}
% \small
% % \resizebox{\columnwidth}{!}
% {
% \label{test-value}
% \scalebox{0.95}{
% \begin{tabular}{lllccc}
% \toprule 
% \begin{tabular}[c]{@{}l@{}} {Poisoning ratio $\rightarrow$}\\{Dataset $\downarrow$}   \end{tabular} & \multicolumn{1}{c}{0.004\%} & \multicolumn{1}{c}{0.01\%} & 0.05\% & 0.1\%   \\ 
% % \hline 
% % \hline
% \midrule
% CIFAR-10 & 8 & 8 & 6 & 6                           \\
% CIFAR-100 & 10 & 8 & 7 & 6                            \\ 
% Tiny ImageNet & 8 & 8 & 7 & 7                          \\ \bottomrule
% \end{tabular}
% }}
% \end{table}
% \vspace{-0.1cm}
\paragraph{\textbf{Evaluation Metrics}}
To evaluate the attack performance of our proposed method, we adopt two metrics: Clean Accuracy (C-Acc), is defined as the prediction accuracy on benign testing samples by the backdoored model, and Attack Success Rate (ASR) measures the accuracy of poisoned testing samples being correctly predicted as the target class. The higher C-Acc and higher ASR imply better attack performance. Furthermore, to assess the performance of detection methods, we utilize three metrics: True Positive Rate (TPR), False Positive Rate (FPR), and Weighted F1 Score ($F_1^{\omega} = \frac{2TP}{2TP+FP+\omega FN'} $) used in ~\cite{yuan2023activation}. TPR reflects the accuracy with which detection methods identify poisoned samples, while FPR indicates the rate at which benign samples are mistakenly classified as poisoned. $F_1^{\omega}$ is one variant of $F_1$ score, which quantifies the detection method's ability to accurately classify poisoned samples while minimizing the misclassification of benign samples.
\begin{table*}[ht]
\caption{The result of 7 compared attack methods and our method under 5 different poisoning ratios in the {CIFAR-10}}
\label{attack_result_CIFAR-10}
\centering
\scriptsize
% \footnotesize
\scalebox{0.705 }{
\resizebox{\textwidth}{!}{
    \begin{tabular}{m{.02\textwidth}m{.11\textwidth} m{.035\textwidth}<{\centering} m{.035\textwidth}<{\centering} m{.035\textwidth}<{\centering} m{.035\textwidth}<{\centering} m{.035\textwidth}<{\centering} m{.035\textwidth}<{\centering} m{0.035\textwidth}<{\centering} m{.035\textwidth}<{\centering} m{.035\textwidth}<{\centering} m{.035\textwidth}<{\centering} m{.035\textwidth}<{\centering} m{.035\textwidth}<{\centering}} %{l cc cc cc cc cc cc}
    \toprule
   { }& {Poisoning ratio $\rightarrow$} & \multicolumn{2}{c}{0.004\%}        & \multicolumn{2}{c}{0.01\%}        & \multicolumn{2}{c}{0.05\%}        & \multicolumn{2}{c}{0.1\%}        & \multicolumn{2}{c}{1\%} 
\\
\cmidrule(lr){3-4} \cmidrule(lr){5-6} \cmidrule(lr){7-8} \cmidrule(lr){9-10} \cmidrule(lr){11-12} \cmidrule(lr){13-14}
  &\multicolumn{1}{c}  {Attack $\downarrow$}  & \multicolumn{1}{c}{C-Acc}    & ASR   & \multicolumn{1}{c}{C-Acc}    & ASR   & \multicolumn{1}{c}{C-Acc}    & ASR   & \multicolumn{1}{c}{C-Acc}    & ASR   & \multicolumn{1}{c}{C-Acc}    & ASR   
\\ 
\hline \hline 
\multirow{8}{*}{\rotatebox{90}{PreAct-ResNet18}} &  BadNets~\cite{gu2019badnets}     
& \multicolumn{1}{c}{93.68} & 0.67 
& \multicolumn{1}{c}{93.74} & 0.88 
& \multicolumn{1}{c}{93.70} & 0.72 
& \multicolumn{1}{c}{93.61} & 1.23 
& \multicolumn{1}{c}{93.14} & 74.73 
\\ 
& Blended~\cite{chen2017targeted}   
& \multicolumn{1}{c}{\textbf{94.21}} & 4.81  
& \multicolumn{1}{c}{94.00} & 7.49  
& \multicolumn{1}{c}{93.81} & 26.07  
& \multicolumn{1}{c}{93.80} & 56.11 
& \multicolumn{1}{c}{93.76} & 94.88   
\\ 
&\scalebox{0.9}{Input-aware}~\cite{nguyen2020input} 
& \multicolumn{1}{c}{91.54} & 12.36  
& \multicolumn{1}{c}{90.28} & 13.73  
& \multicolumn{1}{c}{90.66} & 38.78  
& \multicolumn{1}{c}{91.94} & 47.32 
& \multicolumn{1}{c}{91.74} & 79.18   
\\ 
&LF~\cite{zeng2021rethinking}     
& \multicolumn{1}{c}{93.93} & 0.94  
& \multicolumn{1}{c}{\textbf{93.98}} & 1.22   
& \multicolumn{1}{c}{\textbf{94.06}} & 1.39 
& \multicolumn{1}{c}{93.55} & 12.72  
& \multicolumn{1}{c}{93.56} & 86.46
\\ 
&SIG~\cite{barni2019new}    
& \multicolumn{1}{c}{93.94} & 0.53  
& \multicolumn{1}{c}{93.75} & 1.07 
& \multicolumn{1}{c}{93.93} & 18.76  
& \multicolumn{1}{c}{\textbf{94.05}} & 54.88  
& \multicolumn{1}{c}{93.68} & 87.71 
\\ 
&SSBA~\cite{li2021invisible}      
& \multicolumn{1}{c}{93.81} & 0.71  
& \multicolumn{1}{c}{94.10} & 0.59 
& \multicolumn{1}{c}{93.90} & 1.27 
& \multicolumn{1}{c}{93.89} & 1.62 
& \multicolumn{1}{c}{93.43} & 73.44
\\ 
&WABA~\cite{drager2022backdoor}  
& \multicolumn{1}{c}{93.95} & 5.87  
& \multicolumn{1}{c}{93.76} & 10.62  
& \multicolumn{1}{c}{93.67} & 68.09 
& \multicolumn{1}{c}{93.77} & 86.16
& \multicolumn{1}{c}{93.65} & 99.19 
\\ 

\rowcolor{lightgray!40} & WPDA(Ours)  
& \multicolumn{1}{c}{ {{{94.03}}}} & { {\textbf{98.12}  }}
& \multicolumn{1}{c}{{ {{93.89}}}} & { {\textbf{98.71} }}
& \multicolumn{1}{c}{{ {{93.58}}}} & { {\textbf{99.80}  }}
& \multicolumn{1}{c}{{ {{93.84}}}} & { {\textbf{99.76} }}
& \multicolumn{1}{c}{{ {\textbf{93.86}}}} & { {\textbf{100.00}}}
\\

\hline

\multirow{8}{*}{\rotatebox{90}{VGG19-BN}}  &BadNets~\cite{gu2019badnets}     
& \multicolumn{1}{c}{92.23} & 0.89 
& \multicolumn{1}{c}{91.96} & 0.82 
& \multicolumn{1}{c}{92.23} & 0.94 
& \multicolumn{1}{c}{92.03} & 1.09 
& \multicolumn{1}{c}{91.73} & 74.66 
\\ 
&Blended~\cite{chen2017targeted}   
& \multicolumn{1}{c}{92.25} & 3.41  
& \multicolumn{1}{c}{91.73} & 5.67  
& \multicolumn{1}{c}{92.32} & 22.24  
& \multicolumn{1}{c}{92.16} & 44.10 
& \multicolumn{1}{c}{92.06} & 92.87   
\\ 
&\scalebox{0.9}{Input-aware}~\cite{nguyen2020input} 
& \multicolumn{1}{c}{89.47} & 10.19  
& \multicolumn{1}{c}{89.40} & 11.89  
& \multicolumn{1}{c}{89.62} & 59.82  
& \multicolumn{1}{c}{89.36} & 77.16 
& \multicolumn{1}{c}{89.02} & 67.30   
\\ 
&LF~\cite{zeng2021rethinking}     
& \multicolumn{1}{c}{92.12} & 2.30  
& \multicolumn{1}{c}{\textbf{92.43}} & 0.76   
& \multicolumn{1}{c}{91.94} & 0.97 
& \multicolumn{1}{c}{92.21} & 0.92  
& \multicolumn{1}{c}{91.41} & 1.89 
\\ 
&SIG~\cite{barni2019new}    
& \multicolumn{1}{c}{92.14} & 0.73  
& \multicolumn{1}{c}{91.97} & 2.31 
& \multicolumn{1}{c}{92.20} & 12.47  
& \multicolumn{1}{c}{91.96} & 33.74  
& \multicolumn{1}{c}{92.03} & 91.06 
\\ 
&SSBA~\cite{li2021invisible}      
& \multicolumn{1}{c}{92.26} & 0.98  
& \multicolumn{1}{c}{92.33} & 0.97 
& \multicolumn{1}{c}{\textbf{92.38}} & 1.21 
& \multicolumn{1}{c}{92.36} & 1.44 
& \multicolumn{1}{c}{91.94} & 38.39
\\ 
&WABA~\cite{drager2022backdoor}  
& \multicolumn{1}{c}{92.08} & 3.03  
& \multicolumn{1}{c}{92.02} & 20.74  
& \multicolumn{1}{c}{91.91} & 48.09 
& \multicolumn{1}{c}{91.87} & 79.13 
& \multicolumn{1}{c}{\textbf{92.27}} & 97.93 
\\ 
% \hline 
\rowcolor{lightgray!40}& {WPDA(Ours)}  
& \multicolumn{1}{c}{ {{\textbf{92.49}}}} & { {\textbf{77.50} }}
& \multicolumn{1}{c}{ {92.29}} & { {\textbf{91.79} }}
& \multicolumn{1}{c}{ {92.33}} & { {\textbf{99.23} } }
& \multicolumn{1}{c}{ {\textbf{92.50}}} & { {\textbf{99.64} }}
& \multicolumn{1}{c}{ {{92.01}}} & { {\textbf{99.99}}}
\\
\bottomrule
\end{tabular}
}}
\end{table*}
\subsection{Attack Performance of WPDA}
\label{Attack Performance on WPDA}
\begin{figure}[htbp]
\centering
\vspace{-0.21cm}
\includegraphics[width=8.85cm,height=2.9cm]{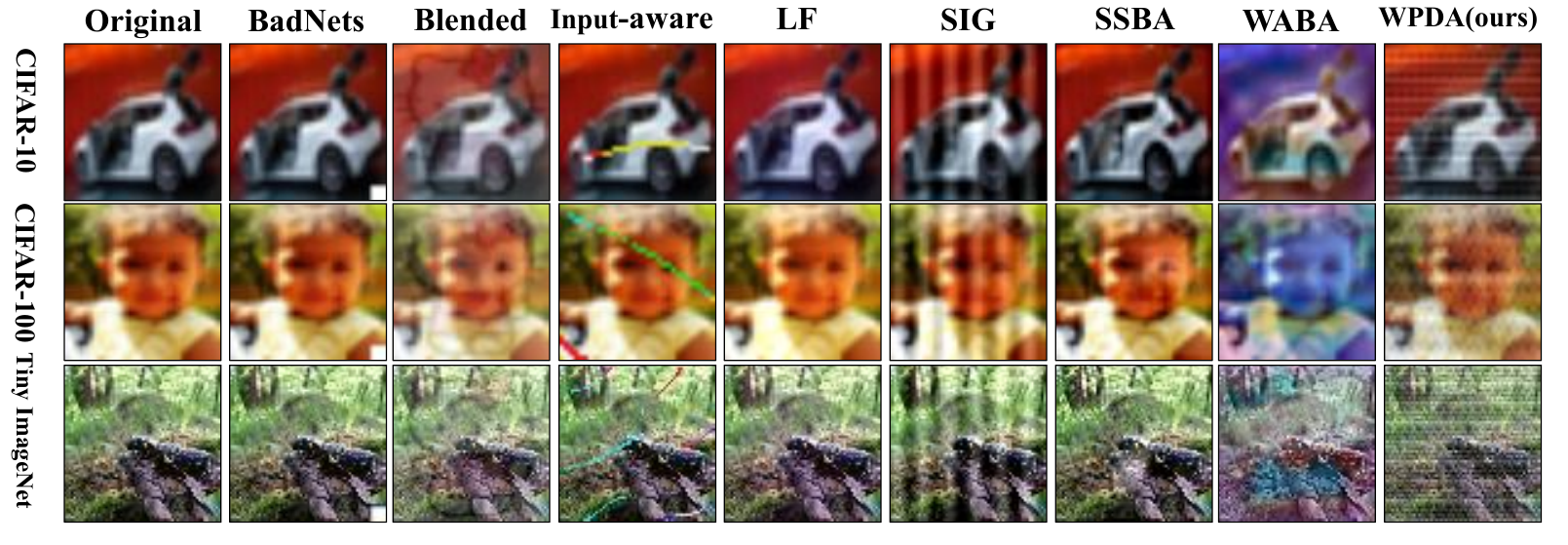}
\caption{Visualization of the poisoned samples in datasets.}
\label{poisoned-samples}
\vspace{-0.5cm}
\end{figure}

In the following, we evaluate the effectiveness performance of 7 compared attack methods and WPDA across 3 datasets and 2 models. 
\paragraph{\textbf{Evaluations on CIFAR-10}}
Tab.~\ref{attack_result_CIFAR-10} illustrates the attack performance of the compared attack methods and our method on CIFAR-10 with various poisoning ratios. It is evident that other methods fail to achieve successful backdoor attacks at the poisoning ratios below 0.01\%. 
At an extremely poisoning ratio of 0.004\%, \textit{i.e.}, only 2 poisoned samples, Input-aware achieves the highest ASR among 7 SOTA methods, reaching 12.36\% on PreAct-ResNet18 and 10.19\% on VGG19-BN. In contrast, WPDA reach an ASR of 98.12\% on PreAct-ResNet18 and 77.50\% on VGG19-BN. As the poisoning ratio increases, all attack methods exhibit a noticeable improvement in performance. However, WPDA stands out by achieving nearly 100\% ASR on both PreAct-ResNet18 and VGG19-BN with a relatively low poisoning ratio of 0.05\%. 
% At an extremely poisoning ratio of $0.004\%$, the highest ASR achieved by Input-aware is $12.36\%$ on PreAct-ResNet18 and $10.19\%$ on VGG19-BN, respectively. In contrast, WPDA reach an ASR of $98.12\%$ on PreAct-ResNet18 and $77.50\%$ on VGG19-BN.
% As the poisoning ratio increases, there is a significant improvement in attack performance. WPDA can achieve ASRs close to $100\%$ on both PreAct-ResNet18 and VGG19-BN with only 0.05\% poisoning ratio. Moreover, the C-Acc of WPDA is the highest among the compared attacks at a poisoning ratio of $0.004\%$, reaching $94.03\%$ on PreAct-ResNet18 and $92.49\%$ on VGG19-BN, respectively. 
\paragraph{\textbf{Evaluations on CIFAR-100}}
Tab.~\ref{attack_result_CIFAR-100} illustrates the performance of backdoor attacks on CIFAR-100,
we notice that the highest ASR achieved by the compared methods is $1.29\%$ on PreAct-ResNet18 and $1.04\%$ on VGG19-BN, which reveals the challenge of successfully embedding a backdoor at an extremely poisoning ratio of $0.004\%$, \textit{i.e.}, only 2 poisoned samples. In contrast, the ASR of WPDA can reach $85.01\%$ on PreAct-ResNet18 and $79.94\%$ on VGG19-BN, respectively. As the poisoning ratio increases, the ASR of most attack methods exhibits an upward trend. While, our method demonstrates a slight decrease in ASR at a poisoning rate of 0.1\% compared to 0.05\%, which is attributed our dynamic adjustment of the trigger intensity $k'$ in poisoned testing samples based on the poisoning ratios, setting a small $k'$ value on a high poisoning ratio. Despite this adjustment causing a slight fluctuation in ASR, our method still achieves the best overall performance compared to the other 7 attack methods. 
\begin{table*}[!ht]
\caption{Results of 7 compared attack methods and our method under 5 different poisoning ratios in the {CIFAR-100}.}
\label{attack_result_CIFAR-100}
\centering
\scriptsize
% \footnotesize
\scalebox{0.705 }{
\resizebox{\textwidth}{!}{
    \begin{tabular}{m{.02\textwidth}m{.11\textwidth} m{.035\textwidth}<{\centering} m{.035\textwidth}<{\centering} m{.035\textwidth}<{\centering} m{.035\textwidth}<{\centering} m{.035\textwidth}<{\centering} m{.035\textwidth}<{\centering} m{0.035\textwidth}<{\centering} m{.035\textwidth}<{\centering} m{.035\textwidth}<{\centering} m{.035\textwidth}<{\centering} m{.035\textwidth}<{\centering} m{.035\textwidth}<{\centering}} %{l cc cc cc cc cc cc}
    \toprule
   { }& {Poisoning ratio $\rightarrow$} & \multicolumn{2}{c}{0.004\%}        & \multicolumn{2}{c}{0.01\%}        & \multicolumn{2}{c}{0.05\%}        & \multicolumn{2}{c}{0.1\%}        & \multicolumn{2}{c}{1\%} 
\\
\cmidrule(lr){3-4} \cmidrule(lr){5-6} \cmidrule(lr){7-8} \cmidrule(lr){9-10} \cmidrule(lr){11-12} \cmidrule(lr){13-14}
  &\multicolumn{1}{c}  {Attack $\downarrow$}  & \multicolumn{1}{c}{C-Acc}    & ASR   & \multicolumn{1}{c}{C-Acc}    & ASR   & \multicolumn{1}{c}{C-Acc}    & ASR   & \multicolumn{1}{c}{C-Acc}    & ASR   & \multicolumn{1}{c}{C-Acc}    & ASR   
\\ 
\hline \hline 
\multirow{8}{*}{\rotatebox{90}{PreAct-ResNet18}} &  BadNets~\cite{gu2019badnets}     
& \multicolumn{1}{c}{\textbf{70.83}} & 0.13 
& \multicolumn{1}{c}{70.23} & 0.17 
& \multicolumn{1}{c}{70.48} & 0.15 
& \multicolumn{1}{c}{70.80} & 0.20 
& \multicolumn{1}{c}{70.25} & 41.51 
\\ 
& Blended~\cite{chen2017targeted}   
& \multicolumn{1}{c}{70.67} & 0.09  
& \multicolumn{1}{c}{\textbf{70.60}} & 0.34  
& \multicolumn{1}{c}{70.74} & 23.37  
& \multicolumn{1}{c}{70.75} & 52.74 
& \multicolumn{1}{c}{\textbf{70.48}} & 90.67    
\\ 
&\scalebox{0.9}{Input-aware}~\cite{nguyen2020input} 
& \multicolumn{1}{c}{65.27} & 1.11  
& \multicolumn{1}{c}{64.88} & 5.47  
& \multicolumn{1}{c}{62.93} & 39.86  
& \multicolumn{1}{c}{64.18} & 73.11 
& \multicolumn{1}{c}{64.82} & 73.72   
\\ 
&LF~\cite{zeng2021rethinking}     
& \multicolumn{1}{c}{70.72} & 0.10  
& \multicolumn{1}{c}{70.42} & 0.09   
& \multicolumn{1}{c}{\textbf{71.02}} & 0.15 
& \multicolumn{1}{c}{\textbf{70.85}} & 0.84  
& \multicolumn{1}{c}{70.36} & 45.55
\\ 
&SIG~\cite{barni2019new}    
& \multicolumn{1}{c}{70.33} & 0.13  
& \multicolumn{1}{c}{70.38} & 0.52 
& \multicolumn{1}{c}{70.75} & 5.04  
& \multicolumn{1}{c}{70.79} & 11.19  
& \multicolumn{1}{c}{69.81} & 77.85 
\\ 
&SSBA~\cite{li2021invisible}      
& \multicolumn{1}{c}{70.32} & 0.05  
& \multicolumn{1}{c}{70.48} & 0.11 
& \multicolumn{1}{c}{70.64} & 0.15 
& \multicolumn{1}{c}{70.31} & 0.40 
& \multicolumn{1}{c}{70.18} & 44.05
\\ 
&WABA~\cite{drager2022backdoor}  
& \multicolumn{1}{c}{70.18} & 1.29  
& \multicolumn{1}{c}{70.37} & 24.46  
& \multicolumn{1}{c}{70.44} & 71.23 
& \multicolumn{1}{c}{70.28} & 86.48
& \multicolumn{1}{c}{70.42} & \textbf{98.56}
\\ 
% \hline 
\rowcolor{lightgray!40}& {WPDA(Ours)}  
& \multicolumn{1}{c}{ {{{70.28}}}} & { {\textbf{85.01} }}
& \multicolumn{1}{c}{{ {{70.56}}}} & { {\textbf{89.01} }}
& \multicolumn{1}{c}{{ {{70.71}}}} & { {\textbf{95.32}  }}
& \multicolumn{1}{c}{{ {{70.44}}}} & { {\textbf{93.22} }}
& \multicolumn{1}{c}{{ {70.32}}} & { {95.74}}
\\
\hline

\multirow{8}{*}{\rotatebox{90}{VGG19-BN}}  &BadNets~\cite{gu2019badnets}     
& \multicolumn{1}{c}{66.06} & 0.15 
& \multicolumn{1}{c}{65.68} & 0.20 
& \multicolumn{1}{c}{65.64} & 0.23 
& \multicolumn{1}{c}{65.62} & 0.32 
& \multicolumn{1}{c}{65.53} & 69.31 
\\ 
&Blended~\cite{chen2017targeted}   
& \multicolumn{1}{c}{65.34} & 0.07  
& \multicolumn{1}{c}{65.25} & 0.08  
& \multicolumn{1}{c}{65.31} & 14.62  
& \multicolumn{1}{c}{65.65} & 37.05 
& \multicolumn{1}{c}{\textbf{65.57}} & 88.12   
\\ 
&\scalebox{0.9}{Input-aware}~\cite{nguyen2020input} 
& \multicolumn{1}{c}{58.01} & 1.04  
& \multicolumn{1}{c}{57.94} & 1.46  
& \multicolumn{1}{c}{60.37} & 1.00  
& \multicolumn{1}{c}{59.87} & 38.35 
& \multicolumn{1}{c}{59.78} & 36.63   
\\ 
&LF~\cite{zeng2021rethinking}     
& \multicolumn{1}{c}{65.47} & 0.17  
& \multicolumn{1}{c}{\textbf{66.11}} & 0.08   
& \multicolumn{1}{c}{65.59} & 0.25 
& \multicolumn{1}{c}{65.61} & 0.27  
& \multicolumn{1}{c}{65.30} & 0.69 
\\ 
&SIG~\cite{barni2019new}    
& \multicolumn{1}{c}{65.45} & 0.31  
& \multicolumn{1}{c}{65.49} & 0.09 
& \multicolumn{1}{c}{\textbf{65.89}} & 4.60  
& \multicolumn{1}{c}{65.34} & 14.73  
& \multicolumn{1}{c}{65.13} & 72.02 
\\ 
&SSBA~\cite{li2021invisible}      
& \multicolumn{1}{c}{65.56} & 0.09  
& \multicolumn{1}{c}{65.84} & 0.08 
& \multicolumn{1}{c}{66.14} & 0.28 
& \multicolumn{1}{c}{\textbf{65.84}} & 0.52 
& \multicolumn{1}{c}{64.87} & 13.03
\\ 
&WABA~\cite{drager2022backdoor}  
& \multicolumn{1}{c}{65.40} & 0.40  
& \multicolumn{1}{c}{65.57} & 7.61  
& \multicolumn{1}{c}{65.48} & 65.72 
& \multicolumn{1}{c}{66.20} & 85.47    
& \multicolumn{1}{c}{65.59} & \textbf{97.64}
\\ 
% \hline 
\rowcolor{lightgray!40}& {WPDA(Ours)}  
& \multicolumn{1}{c}
 { {\textbf{66.24}}} & 
{ {\textbf{79.94}  }}
& \multicolumn{1}{c}{ {65.18}} & 
{ {\textbf{88.09} }}
& \multicolumn{1}{c}{ {65.62}} & 
{ {\textbf{97.20}  } }
& \multicolumn{1}{c}{ {65.47}} & 
{ {\textbf{91.06}  }}
& \multicolumn{1}{c}{ {{65.13}}} & 
{ {{94.58}}}  
\\
\bottomrule
\end{tabular}
}}
\end{table*}
\paragraph{\textbf{Evaluations on Tiny ImageNet}}
The attack performance on Tiny ImageNet is shown in Tab.~\ref{attack_result_tiny}. Due to the limited number of 50 samples per class in Tiny ImageNet, SIG, as a clean-label attack, can only implement backdoor attacks at poisoning ratios below 1\%. At an extremely poisoning ratio of 0.004\%, \textit{i.e.}, only 4 poisoned samples, WPDA reaches a ASR of 85.01\% on PreAct-ResNet18 and 79.94\% on VGG19-BN, significantly outperforming the compared methods. To achieve the same attack performance, the compared methods require a higher poisoning ratio. For instance, WABA needs at least a 0.1\% poisoning ratio, SIG and blended require at least 1\%, and other methods need even much higher poisoning ratios.
\begin{table*}[!ht]
\caption{Results of 7 compared attack methods and our method under 5 different poisoning ratios in the {Tiny ImageNet}.}
\label{attack_result_tiny}
\centering
\scriptsize
% \footnotesize
\scalebox{0.705 }{
\resizebox{\textwidth}{!}{
    \begin{tabular}{m{.02\textwidth}m{.11\textwidth} m{.035\textwidth}<{\centering} m{.035\textwidth}<{\centering} m{.035\textwidth}<{\centering} m{.035\textwidth}<{\centering} m{.035\textwidth}<{\centering} m{.035\textwidth}<{\centering} m{0.035\textwidth}<{\centering} m{.035\textwidth}<{\centering} m{.035\textwidth}<{\centering} m{.035\textwidth}<{\centering} m{.035\textwidth}<{\centering} m{.035\textwidth}<{\centering}} %{l cc cc cc cc cc cc}
    \toprule
   { }& {Poisoning ratio $\rightarrow$} & \multicolumn{2}{c}{0.004\%}        & \multicolumn{2}{c}{0.01\%}        & \multicolumn{2}{c}{0.05\%}        & \multicolumn{2}{c}{0.1\%}        & \multicolumn{2}{c}{1\%} 
\\
\cmidrule(lr){3-4} \cmidrule(lr){5-6} \cmidrule(lr){7-8} \cmidrule(lr){9-10} \cmidrule(lr){11-12} \cmidrule(lr){13-14}
  &\multicolumn{1}{c}  {Attack $\downarrow$}  & \multicolumn{1}{c}{C-Acc}    & ASR   & \multicolumn{1}{c}{C-Acc}    & ASR   & \multicolumn{1}{c}{C-Acc}    & ASR   & \multicolumn{1}{c}{C-Acc}    & ASR   & \multicolumn{1}{c}{C-Acc}    & ASR   
\\ 
\hline \hline 
\multirow{8}{*}{\rotatebox{90}{PreAct-ResNet18}} &  BadNets~\cite{gu2019badnets}     
& \multicolumn{1}{c}{56.87} & 0.12 
& \multicolumn{1}{c}{56.86} & 0.18 
& \multicolumn{1}{c}{57.26} & 0.25 
& \multicolumn{1}{c}{57.51} & 33.53 
& \multicolumn{1}{c}{57.51} & 95.01 
\\ 
& Blended~\cite{chen2017targeted}   
& \multicolumn{1}{c}{57.24} & 0.06  
& \multicolumn{1}{c}{57.11} & 2.90  
& \multicolumn{1}{c}{57.68} & 50.88  
& \multicolumn{1}{c}{57.58} & 65.19 
& \multicolumn{1}{c}{{57.31}} & 95.29   
\\ 
&\scalebox{0.9}{Input-aware}~\cite{nguyen2020input} 
& \multicolumn{1}{c}{\textbf{58.55}} & 2.00   
& \multicolumn{1}{c}{\textbf{57.93}} & 34.96  
& \multicolumn{1}{c}{\textbf{58.05}} & 57.29  
& \multicolumn{1}{c}{\textbf{57.96}} & 63.80 
& \multicolumn{1}{c}{\textbf{57.94}} & 85.12   
\\ 
&LF~\cite{zeng2021rethinking}     
& \multicolumn{1}{c}{56.98} & 0.17  
& \multicolumn{1}{c}{57.03} & 0.13   
& \multicolumn{1}{c}{{57.26}} & 0.16 
& \multicolumn{1}{c}{{56.97}} & 19.49  
& \multicolumn{1}{c}{56.47} & 83.03
\\ 
&SIG~\cite{barni2019new}    
& \multicolumn{1}{c}{57.11} & 4.14  
& \multicolumn{1}{c}{57.16} & 10.32 
& \multicolumn{1}{c}{56.86} & 27.42  
& \multicolumn{1}{c}{57.03} & 44.93  
& \multicolumn{1}{c}{---} & --- 
\\ 
&SSBA~\cite{li2021invisible}      
& \multicolumn{1}{c}{56.80} & 0.11  
& \multicolumn{1}{c}{56.96} & 0.30 
& \multicolumn{1}{c}{57.23} & 2.31 
& \multicolumn{1}{c}{57.20} & 13.28 
& \multicolumn{1}{c}{56.88} & 80.43
\\ 
&WABA~\cite{drager2022backdoor}  
& \multicolumn{1}{c}{57.10} & 4.13  
& \multicolumn{1}{c}{57.30} & 15.92  
& \multicolumn{1}{c}{56.51} & 74.14
& \multicolumn{1}{c}{57.17} & 90.13
& \multicolumn{1}{c}{56.82} & {98.37}
\\ 
% \hline 
\rowcolor{lightgray!40}& {WPDA(Ours)}  
& \multicolumn{1}{c}{{{57.24}}} & {{\textbf{80.99} }}
& \multicolumn{1}{c}{{{{56.88}}}} & {{\textbf{85.80}  }}
& \multicolumn{1}{c}{{{{56.73}}}} & {{\textbf{88.45} }}
& \multicolumn{1}{c}{{{{56.86}}}} & {{\textbf{90.71}}}
& \multicolumn{1}{c}{{{57.28}}} & {{\textbf{98.53}}}
\\
\hline

\multirow{8}{*}{\rotatebox{90}{VGG19-BN}}  &BadNets~\cite{gu2019badnets}     
& \multicolumn{1}{c}{53.65} & 0.24 
& \multicolumn{1}{c}{53.99} & 0.18 
& \multicolumn{1}{c}{53.87} & 0.38 
& \multicolumn{1}{c}{54.85} & 48.85 
& \multicolumn{1}{c}{53.48} & \textbf{98.86}
\\ 
&Blended~\cite{chen2017targeted}   
& \multicolumn{1}{c}{53.77} & 0.23 
& \multicolumn{1}{c}{53.71} & 5.88  
& \multicolumn{1}{c}{53.53} & 52.88  
& \multicolumn{1}{c}{53.43} & 69.38 
& \multicolumn{1}{c}{{52.47}} & 94.90   
\\ 
&\scalebox{0.9}{Input-aware}~\cite{nguyen2020input} 
& \multicolumn{1}{c}{53.80} & 8.91   
& \multicolumn{1}{c}{53.98} & 17.71  
& \multicolumn{1}{c}{53.92} & 35.13  
& \multicolumn{1}{c}{\textbf{54.09}} & 85.43 
& \multicolumn{1}{c}{\textbf{54.10}} & 79.82   
\\ 
&LF~\cite{zeng2021rethinking}     
& \multicolumn{1}{c}{54.02} & 0.16  
& \multicolumn{1}{c}{53.28} & 0.17   
& \multicolumn{1}{c}{{53.25}} & 0.22 
& \multicolumn{1}{c}{{52.80}} & 0.22  
& \multicolumn{1}{c}{51.94} & 0.51
\\ 
&SIG~\cite{barni2019new}    
& \multicolumn{1}{c}{53.35} & 3.09 
& \multicolumn{1}{c}{54.24} & 1.65 
& \multicolumn{1}{c}{53.62} & 12.41  
& \multicolumn{1}{c}{53.33} & 26.92  
& \multicolumn{1}{c}{---} & --- 
\\ 
&SSBA~\cite{li2021invisible}      
& \multicolumn{1}{c}{\textbf{54.15}} & 0.02  
& \multicolumn{1}{c}{53.17} & 0.22 
& \multicolumn{1}{c}{\textbf{54.35}} & 2.42 
& \multicolumn{1}{c}{53.91} & 8.42 
& \multicolumn{1}{c}{53.10} & 86.63
\\ 
&WABA~\cite{drager2022backdoor}  
& \multicolumn{1}{c}{52.43} & 9.76  
& \multicolumn{1}{c}{53.20} & 24.90  
& \multicolumn{1}{c}{53.80} & 80.37
& \multicolumn{1}{c}{53.80} & 93.40
& \multicolumn{1}{c}{{53.22}} & {98.50}
\\ 
% \hline 
\rowcolor{lightgray!40}& {WPDA(Ours)}  
& \multicolumn{1}{c}{{{{53.46}}}} & {{\textbf{64.96} }}
& \multicolumn{1}{c}{{\textbf{54.25}}} & {{\textbf{81.89}}}
& \multicolumn{1}{c}{{53.74}} & {{\textbf{91.96} } }
& \multicolumn{1}{c}{{53.99}} & {{\textbf{90.30} }}
& \multicolumn{1}{c}{{{51.39}}} & {{{96.03}}}
\\
\bottomrule
\end{tabular}
}}
\end{table*}
\paragraph{\textbf{Summary}}
Our method demonstrates superiority over 7 other attack methods across three datasets and two models, exhibiting excellent attack effectiveness, particularly at an extremely low poisoning ratio of 0.004\%, \textit{i.e.}, only 2 poisoned training samples in CIFAR-10 and CIFAR-100, and 4 poisoned training samples in Tiny ImageNet.
Besides, poisoned testing samples generated by WPDA maintain stealthiness in visual (shown in Fig.~\ref{poisoned-samples}).
% At extremely low poisoning ratios, especially blow WPDA achieves
% Besides, Backdoor attacks under the low poisoning ratios have a minimal impact on clean accuracy. 
% The performance of other attack methods underscores the significant challenges associated with backdoor attacks under low poisoning ratios. Our method strategically selects the poisoning frequency regions based on the characteristics of the dataset, enhancing the efficiency of the attack and achieving a high ASR even with only 2 poisoned training samples on CIFAR-10 and CIFAR-100, 4 poisoned training samples on Tiny ImageNet. Besides, poisoned testing samples of WPDA maintain the visual naturalness (shown in Fig.~\ref{poisoned-samples}).
% The above comparisons demonstrate the superior attack performance of WPDA compared to 7 SOTA attacks, even at low poisoning ratios.
% Fig.~\ref{poisoned-samples} shows the visualization of the poisoned samples on CIFAR-10, CIFAR-100 and Tiny ImageNet. 
% Compared with other attack methods. 
% Although the backdoor samples based on WPDA method is not the most natural to the original image in visual, WPDA balances the effectiveness of the backdoor attack and the naturalness of the samples at a very low poisoning ratio.
% \vspace{-0.3cm}
\subsection{Detection Performance of WPDA}
\label{Detection Performance on WPDA}
In the following, we evaluate the stealthiness performance of WPDA. Considering that 7 compared attack methods compared at low poisoning ratios fail to achieve successful backdoor attacks, so we only evaluate the performance of WPDA on 3 detection methods. The results on VGG19-BN are as follows, and the results on PreAct-ResNet18 are 
% similar to that on VGG19-BN, and the details on PreAct-ResNet18 are 
shown in Supplementary
% ~\ref{detection on pre}.

\begin{table*}[!ht]
\caption{The result of WPDA against 3 detection methods under 5 different poisoning ratios in the datasets. The training model adopts VGG19-BN. }
\label{detection_result_vgg}
\centering
% \small
\normalsize
\scalebox{0.82}{
\resizebox{\textwidth}{!}{
    \begin{tabular}{m{.11\textwidth} m{.1\textwidth} m{.0355\textwidth}<{\centering} m{.0355\textwidth}<{\centering} m{.0355\textwidth}<{\centering} m{.0355\textwidth}<{\centering} m{.0355\textwidth}<{\centering} m{.0355\textwidth}<{\centering} m{.0355\textwidth}<{\centering} m{.0355\textwidth}<{\centering} m{.0355\textwidth}<{\centering} m{.0355\textwidth}<{\centering} m{.0355\textwidth}<{\centering} m{.0355\textwidth}<{\centering} m{.0355\textwidth}<{\centering} m{.0355\textwidth}<{\centering} m{.0355\textwidth}<{\centering} m{.0355\textwidth}<{\centering} m{.0355\textwidth}<{\centering} m{.0355\textwidth}<{\centering}} %{l l cc cc cc cc cc cc cc cc cc}
    \toprule
  \multirow{2}{*}{Dataset} & Poisoning ratio $\rightarrow$ & \multicolumn{3}{c}{0.004\%}        & \multicolumn{3}{c}{0.01\%}        & \multicolumn{3}{c}{0.05\%}        & \multicolumn{3}{c}{0.1\%}        & \multicolumn{3}{c}{1\%} 
\\
\cmidrule(lr){3-5} \cmidrule(lr){6-8} \cmidrule(lr){9-11} \cmidrule(lr){12-14} \cmidrule(lr){15-17} \cmidrule(lr){18-20}
 & Detection $\downarrow$  & \multicolumn{1}{c}{TPR }    & FPR    & $F^{\omega}_1$    & \multicolumn{1}{c}{TPR }    & FPR    & $F^{\omega}_1$    & \multicolumn{1}{c}{TPR }    & FPR    & $F^{\omega}_1$    & \multicolumn{1}{c}{TPR }    & FPR    & $F^{\omega}_1$    & \multicolumn{1}{c}{TPR }    & FPR    & $F^{\omega}_1$       
\\ 
\hline \hline 
% \multirow{4}{*}{CIFAR10} & AC~\cite{chen2018detecting}        
% & \multicolumn{1}{c}{0.00} & 0.08 & 0.00
% & \multicolumn{1}{c}{0.00} & 0.11 & 0.00
% & \multicolumn{1}{c}{0.00} & 0.11 & 0.00
% & \multicolumn{1}{c}{0.00} & 0.11 & 0.00
% & \multicolumn{1}{c}{0.99} & 0.05 & 0.27
% \\ 
%   & SCAn~\cite{tang2021demon}   
% & \multicolumn{1}{c}{0.00} & 0.00 & 0.00 
% & \multicolumn{1}{c}{0.00} & 0.00 & 0.00  
% & \multicolumn{1}{c}{0.00} & 0.00 & 0.00  
% & \multicolumn{1}{c}{0.00} & 0.05 & 0.00
% & \multicolumn{1}{c}{0.98} & 0.00 & 0.93  
% \\ 
%   & Spectral~\cite{tran2018spectral}  
% & \multicolumn{1}{c}{0.00} & 0.15 & 0.00 
% & \multicolumn{1}{c}{0.00} & 0.15 & 0.00 
% & \multicolumn{1}{c}{0.00} & 0.15 & 0.00  
% & \multicolumn{1}{c}{0.00} & 0.15 & 0.00  
% & \multicolumn{1}{c}{1.00} & 0.14 & 0.12
% \\ 
%   & STRIP~\cite{gao2019strip}    
% & \multicolumn{1}{c}{0.00} & 0.10 & 0.00 
% & \multicolumn{1}{c}{0.00} & 0.14 & 0.00 
% & \multicolumn{1}{c}{0.32} & 0.07 & 0.00  
% & \multicolumn{1}{c}{0.42} & 0.09 & 0.01  
% & \multicolumn{1}{c}{0.95} & 0.13 & 0.13
\multirow{4}{*}{CIFAR10} & AC~\cite{chen2018detecting}        
& \multicolumn{1}{c}{0.00} & 8.42 & 0.00
& \multicolumn{1}{c}{0.00} & 10.92 & 0.00
& \multicolumn{1}{c}{0.00} & 11.31 & 0.00
& \multicolumn{1}{c}{0.00} & 11.31 & 0.00
& \multicolumn{1}{c}{99.40} & 7.41 & 27.17
\\ 
  & SCAn~\cite{tang2021demon}   
& \multicolumn{1}{c}{0.00} & 0.00 & 0.00 
& \multicolumn{1}{c}{0.00} & 0.00 & 0.00  
& \multicolumn{1}{c}{0.00} & 0.00 & 0.00  
& \multicolumn{1}{c}{0.00} & 4.58 & 0.00
& \multicolumn{1}{c}{98.40} & 0.00 & 93.18  
% \\ 
%   & Spectral~\cite{tran2018spectral}  
% & \multicolumn{1}{c}{0.00} & 15.00 & 0.00 
% & \multicolumn{1}{c}{0.00} & 15.00 & 0.00 
% & \multicolumn{1}{c}{0.00} & 15.01 & 0.00  
% & \multicolumn{1}{c}{0.00} & 15.02 & 0.00  
% & \multicolumn{1}{c}{100.00} & 29.30 & 6.45
\\ 
  & STRIP~\cite{gao2019strip}    
& \multicolumn{1}{c}{0.00} & 10.02 & 0.00 
& \multicolumn{1}{c}{0.00} & 13.79 & 0.00 
& \multicolumn{1}{c}{32.00} & 6.64 & 0.46 
& \multicolumn{1}{c}{42.00} & 8.99 & 0.88  
& \multicolumn{1}{c}{94.80} & 12.70 & 12.69
\\
\hline
% \multirow{4}{*}{CIFAR100} & AC~\cite{chen2018detecting}     
% & \multicolumn{1}{c}{0.00} & 0.04 & 0.00
% & \multicolumn{1}{c}{0.00} & 0.03 & 0.00
% & \multicolumn{1}{c}{0.00} & 0.01 & 0.00
% & \multicolumn{1}{c}{0.00} & 0.03 & 0.00
% & \multicolumn{1}{c}{0.00} & 0.02 & 0.00
% \\ 
%   & SCAn~\cite{tang2021demon} 
% & \multicolumn{1}{c}{0.00} & 0.00 & 0.00 
% & \multicolumn{1}{c}{0.00} & 0.00 & 0.00  
% & \multicolumn{1}{c}{0.00} & 0.00 & 0.00  
% & \multicolumn{1}{c}{0.00} & 0.05 & 0.00
% & \multicolumn{1}{c}{0.98} & 0.00 & 0.93  
% \\ 
%   & Spectral~\cite{tran2018spectral}  
% & \multicolumn{1}{c}{0.00} & 0.15 & 0.00 
% & \multicolumn{1}{c}{0.00} & 0.15 & 0.00 
% & \multicolumn{1}{c}{0.00} & 0.15 & 0.00  
% & \multicolumn{1}{c}{0.40} & 0.15 & 0.00  
% & \multicolumn{1}{c}{0.30} & 0.15 & 0.03
% \\ 
%   & STRIP~\cite{gao2019strip}    
% & \multicolumn{1}{c}{0.00} & 0.13 & 0.14 
% & \multicolumn{1}{c}{0.00} & 0.12 & 0.00 
% & \multicolumn{1}{c}{0.32} & 0.19 & 0.00  
% & \multicolumn{1}{c}{0.42} & 0.17 & 0.01  
% & \multicolumn{1}{c}{0.95} & 0.15 & 0.12
\multirow{4}{*}{CIFAR100} & AC~\cite{chen2018detecting}     
& \multicolumn{1}{c}{0.00} & 3.67 & 0.00
& \multicolumn{1}{c}{0.00} & 3.03 & 0.00
& \multicolumn{1}{c}{0.00} & 1.24 & 0.00
& \multicolumn{1}{c}{0.00} & 2.57 & 0.00
& \multicolumn{1}{c}{0.00} & 2.03 & 0.00
\\ 
  & SCAn~\cite{tang2021demon} 
& \multicolumn{1}{c}{0.00} & 0.00 & 0.00 
& \multicolumn{1}{c}{0.00} & 0.00 & 0.00  
& \multicolumn{1}{c}{0.00} & 0.00 & 0.00  
& \multicolumn{1}{c}{0.00} & 0.00 & 0.00
& \multicolumn{1}{c}{0.00} & 0.00 & 0.00  
% \\ 
%   & Spectral~\cite{tran2018spectral}  
% & \multicolumn{1}{c}{0.00} & 15.00 & 0.00 
% & \multicolumn{1}{c}{0.00} & 15.00 & 0.00 
% & \multicolumn{1}{c}{0.00} & 15.02 & 0.00  
% & \multicolumn{1}{c}{4.00} & 15.03 & 0.05  
% & \multicolumn{1}{c}{29.8} & 14.91 & 2.75
\\ 
  & STRIP~\cite{gao2019strip}    
& \multicolumn{1}{c}{0.00} & 13.65 & 0.00
& \multicolumn{1}{c}{40.00} & 11.54 & 0.07
& \multicolumn{1}{c}{52.00} & 19.00 & 0.27  
& \multicolumn{1}{c}{70.00} & 16.52 & 0.83  
& \multicolumn{1}{c}{100.00} & 15.10 & 11.80
% & \multicolumn{1}{c}{0.00} & 13.67 & 0.00
\\
\hline
% \multirow{4}{*}{Tiny ImageNet} & AC~\cite{chen2018detecting}     
% & \multicolumn{1}{c}{0.00} & 0.03 & 0.00
% & \multicolumn{1}{c}{0.00} & 0.06 & 0.00
% & \multicolumn{1}{c}{0.00} & 0.07 & 0.00
% & \multicolumn{1}{c}{0.00} & 0.08 & 0.00
% & \multicolumn{1}{c}{0.00} & 0.18 & 0.00
% \\ 
%   & SCAn~\cite{tang2021demon} 
% & \multicolumn{1}{c}{0.00} & 0.00 & 0.00 
% & \multicolumn{1}{c}{0.00} & 0.00 & 0.00  
% & \multicolumn{1}{c}{0.00} & 0.00 & 0.00  
% & \multicolumn{1}{c}{0.00} & 0.00 & 0.00
% & \multicolumn{1}{c}{0.99} & 0.00 & 0.96 
% \\ 
%   & Spectral~\cite{tran2018spectral}  
% & \multicolumn{1}{c}{0.00} & 0.15 & 0.00 
% & \multicolumn{1}{c}{0.00} & 0.15 & 0.00 
% & \multicolumn{1}{c}{0.28} & 0.15 & 0.00  
% & \multicolumn{1}{c}{0.00} & 0.15 & 0.00  
% & \multicolumn{1}{c}{0.17} & 0.15 & 0.01
% \\ 
%   & STRIP~\cite{gao2019strip}    
% & \multicolumn{1}{c}{0.00} & 0.17 & 0.00 
% & \multicolumn{1}{c}{0.00} & 0.16 & 0.00 
% & \multicolumn{1}{c}{0.10} & 0.13 & 0.00  
% & \multicolumn{1}{c}{0.09} & 0.14 & 0.00  
% & \multicolumn{1}{c}{0.47} & 0.16 & 0.04
\multirow{4}{*}{Tiny ImageNet} & AC~\cite{chen2018detecting}     
& \multicolumn{1}{c}{0.00} & 2.98 & 0.00
& \multicolumn{1}{c}{0.00} & 6.26 & 0.00
& \multicolumn{1}{c}{0.00} & 7.13 & 0.00
& \multicolumn{1}{c}{0.00} & 7.98 & 0.00
& \multicolumn{1}{c}{0.00} & 18.47 & 0.00
\\ 
  & SCAn~\cite{tang2021demon} 
& \multicolumn{1}{c}{0.00} & 0.00 & 0.00 
& \multicolumn{1}{c}{0.00} & 0.00 & 0.00  
& \multicolumn{1}{c}{0.00} & 0.00 & 0.00  
& \multicolumn{1}{c}{0.00} & 0.00 & 0.00
& \multicolumn{1}{c}{99.10} & 0.00 & 96.03 
% \\ 
%   & Spectral~\cite{tran2018spectral}  
% & \multicolumn{1}{c}{0.00} & 15.00 & 0.00 
% & \multicolumn{1}{c}{0.00} & 15.00 & 0.00 
% & \multicolumn{1}{c}{28.00} & 15.00 & 0.18  
% & \multicolumn{1}{c}{0.00} & 15.03 & 0.00  
% & \multicolumn{1}{c}{17.00} & 15.05 & 1.50
\\ 
  & STRIP~\cite{gao2019strip}    
& \multicolumn{1}{c}{0.00} & 16.70 & 0.00 
& \multicolumn{1}{c}{0.00} & 16.27 & 0.00 
& \multicolumn{1}{c}{10.00} & 12.88 & 0.08  
& \multicolumn{1}{c}{9.00} & 13.74 & 0.12  
& \multicolumn{1}{c}{47.20} & 15.58 & 4.47
\\
\bottomrule
\end{tabular}
}}
\end{table*}

\begin{figure}[htbp]
\centering
\vspace{-0.31cm}
\includegraphics[width=8.82cm,height=2.9cm]
{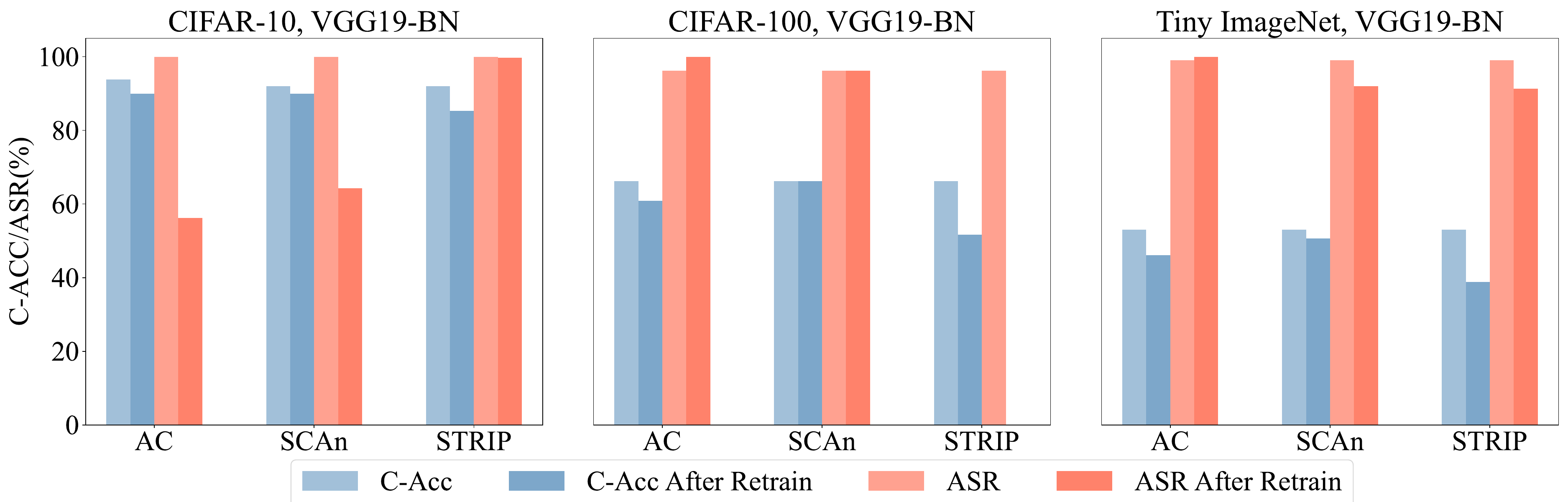}
% {figures/detection_retrain_vgg.pdf}
\caption{Results of retraining after removing suspicious samples using detection methods at 1\% poisoning ratio on VGG19-BN.}
\label{detection vgg}
\vspace{-0.3cm}
\end{figure}

\paragraph{\textbf{Evaluations on AC}}
The core principle of AC lies in the observation that poisoned training samples activate specific neurons that are different from those activated by benign training samples. By leveraging activation clustering, samples are divided into two clusters in each class, and the smaller one is regarded as poisoned. For low poisoning ratios, Tab.~\ref{detection_result_vgg} indicates that AC fails to correctly identify poisoned training samples, which is because the number of poisoned training samples is too small to be recognized as a distinct cluster by AC. Meanwhile, the activations of some benign samples are incorrectly divided into two clusters, with the smaller one being misidentified as poisoned. For 1\% poisoning ratio, AC shows significant differences in detection performance across datasets. On CIFAR-10, the TPR reaches 99.4\%, while it is only 0.00\% on CIFAR-100 and Tiny ImageNet. This difference is primarily due to the number of samples per class: 5,000 samples per class in CIFAR-10, whereas there are 500 samples per class in CIFAR-100 and 50 samples per class in Tiny ImageNet. With 500 poisoned training samples, AC can more easily cluster the samples in CIFAR-10 into two clusters and accurately identify the smaller one as poisoned.  In contrast, in CIFAR-100 and Tiny ImageNet, the large number of poisoned training samples makes it easier for AC to misidentify the larger cluster with benign samples as poisoned, resulting in a high FPR, such as a FPR of 18.47\% in Tiny ImageNet. In CIFAR-10, although 497 out of 500 poisoned samples are detected, 3 poisoned samples still remain. Fig.~\ref{detection vgg} demonstrates that a successful backdoor attack can still be achieved using these remaining 3 samples. The difference in attack effectiveness compared to 0.004\% poisoning ratio is due to the small $k'$ in the poisoned testing samples at 1\% poisoning ratio. The above results indicate that WPDA can evade the detection of AC effectively, particularly at low poisoning ratios.
\paragraph{\textbf{Evaluations on SCAn}}
SCAn leverages a statistical analysis method based on feature representations, identifying poisoned classes and poisoned training samples by examining whether their distribution aligns with a mixture of normal distributions. In Tab.~\ref{detection_result_vgg}, for low poisoning ratios, the small number of poisoned training samples poses a challenge for SCAn. With a negligible statistical footprint, these samples become difficult to identify through statistical analysis methods, hindering the effectiveness of SCAn. For 1\% poisoning ratio, SCAn fails to detect the poisoned training samples in the CIFAR-100. However, due to differences among the datasets, some poisoned training samples are identified in CIFAR-10 and Tiny ImageNet. Even if 492 out of 500 poisoned training samples in CIFAR-10 and 991 out of 1,000 poisoned training samples in Tiny ImageNet are removed, the remaining samples still achieve successful backdoor attacks, as evaluated in Fig.~\ref{detection vgg}. These results demonstrate the stealthiness of WPDA under the detection of SCAn.

\paragraph{\textbf{Evaluations on STRIP}}
STRIP is a detection method based on the perturbation of inputs. The principle behind this detection method is that the predictions on perturbed poisoned samples are more consistent than the predictions on benign samples. Tab.~\ref{detection_result_vgg} demonstrates a positive correlation between the poisoning ratios and TPR. For low poisoning ratios, STRIP fails to detect all poisoned samples and misidentify lots of benign samples as poisoned. This not only fails to effectively resist WPDA backdoor attacks, but also degrades the model's performance on benign samples. For 1\% poisoning ratio, on CIFAR-100, STRIP successfully detects all poisoned samples. However, Fig.~\ref{detection vgg} demonstrates that the model's accuracy on benign samples significantly decreases due to STRIP misidentifying a large number of benign samples as poisoned, indicating that it is difficult for STRIP to resist the WPDA. On CIFAR-10 and Tiny ImageNet, the poisoned samples that undetected by STRIP still retain a strong backdoor effect, achieving a high ASR. These above results indicate STRIP cannot resist WPDA successfully.

\paragraph{\textbf{Summary}}
Above results reveal the challenge existing detection methods face under low poisoning ratios, which struggle to accurately identify all poisoned training samples. Specifically at 0.004\% poisoning ratio, AC, SCAn, and STRIP failed to detect any poisoned samples generated by WPDA across three datasets, demonstrating the stealthiness of WPDA.
%%%%%%%%%%%%%%%%%%%%%%%%%%%%%%%%%%%%%%%%%%%%%%%%%%%%%%%%%%%%%%%%%%%%%%%%%%%%%%%%%%%%%%%%%%%%%%%%%%%%%%%%%%

\subsection{Defense Performance of WPDA}
\label{Defense Performance on WPDA}
\begin{figure}[t]
\centering  %图片全局居中
    \subfigure[CIFAR-10]{ 
    \includegraphics[width=8.2cm,height =4.0cm]
    % {figures/cifar10_vgg19_bn_1_.pdf}
    {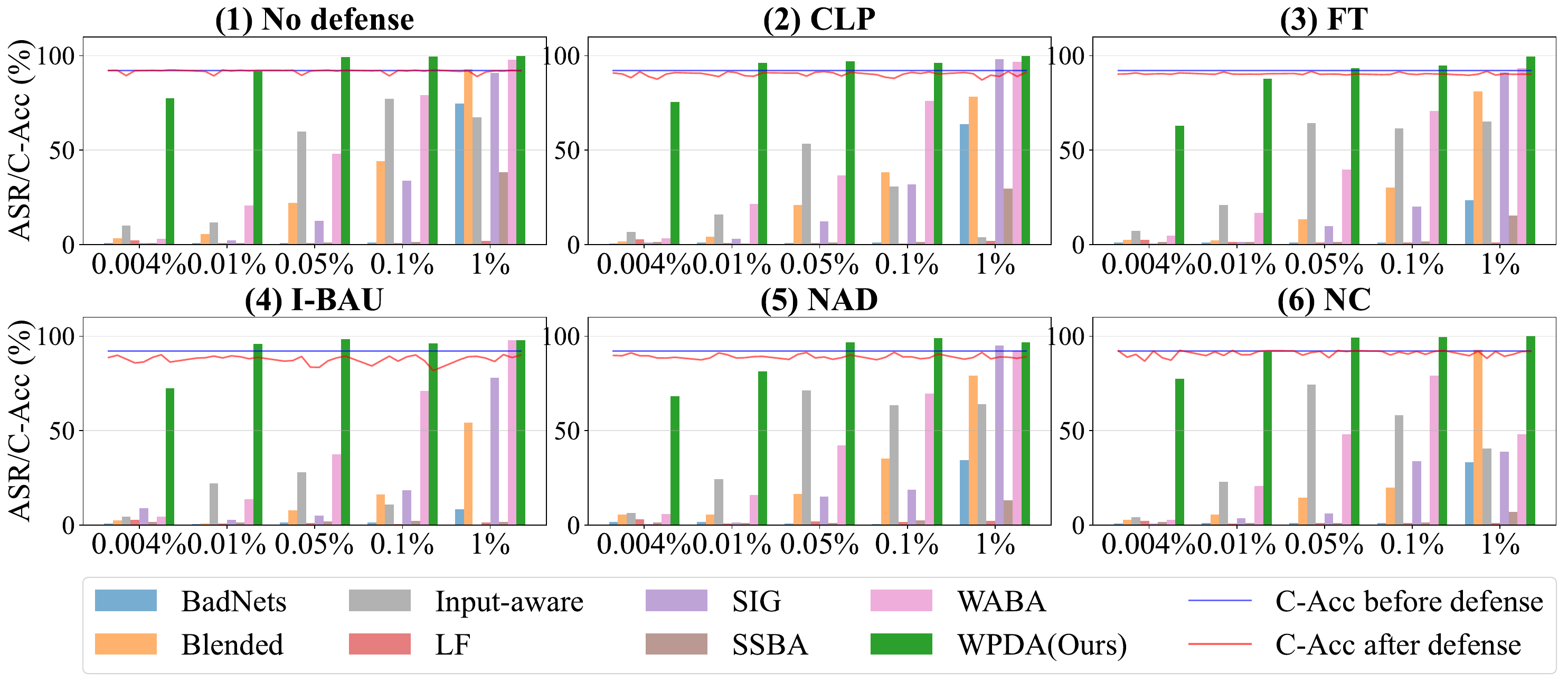} 
    \label{CIFAR-10_vgg_defense}}
    \subfigure[CIFAR-100]{
    \includegraphics[width=8.2cm,height =4.0cm]
    % {figures/cifar100_vgg19_bn_1_.pdf}
    {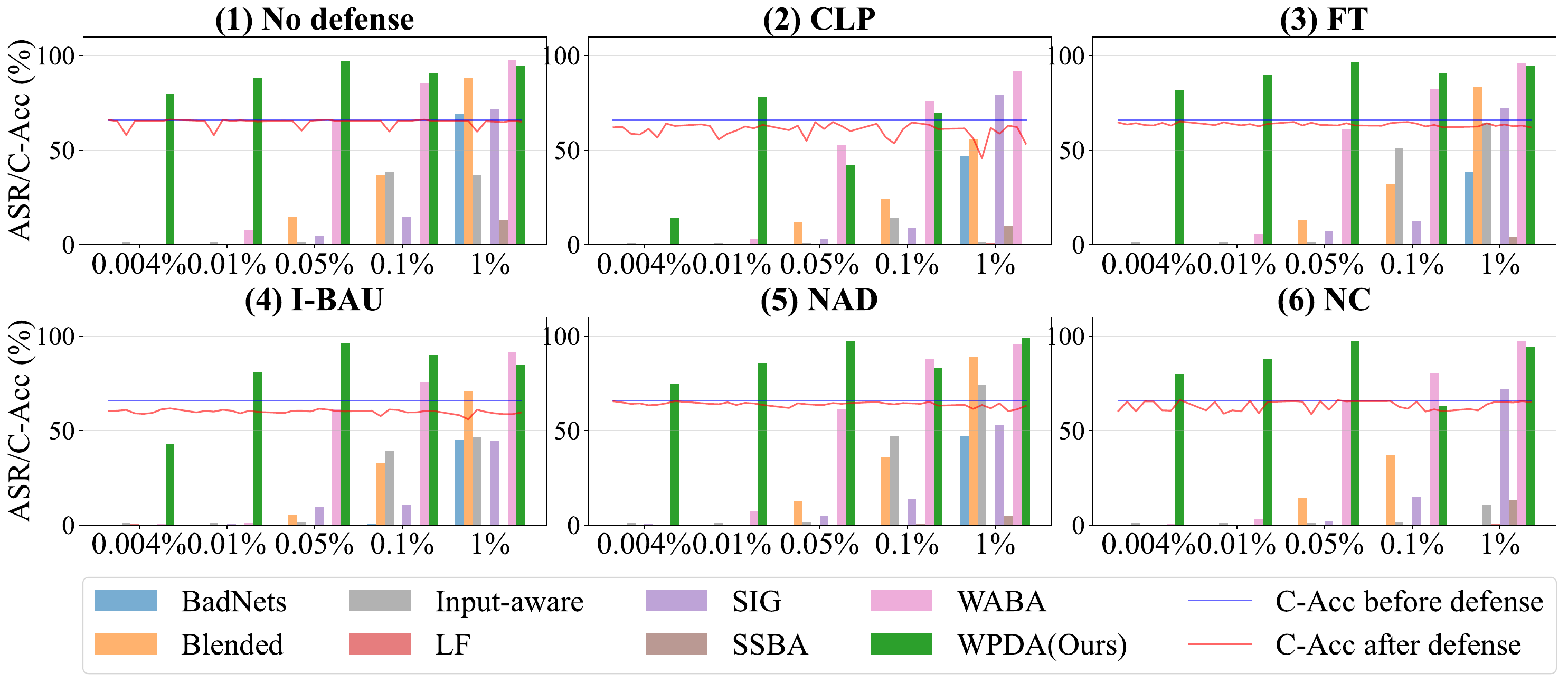}
    \label{CIFAR-100_vgg_defense}}
    \subfigure[Tiny ImageNet]{
    \includegraphics[width=8.2cm,height =4.0cm]
    % {figures/tiny_vgg19_bn_.pdf}
    {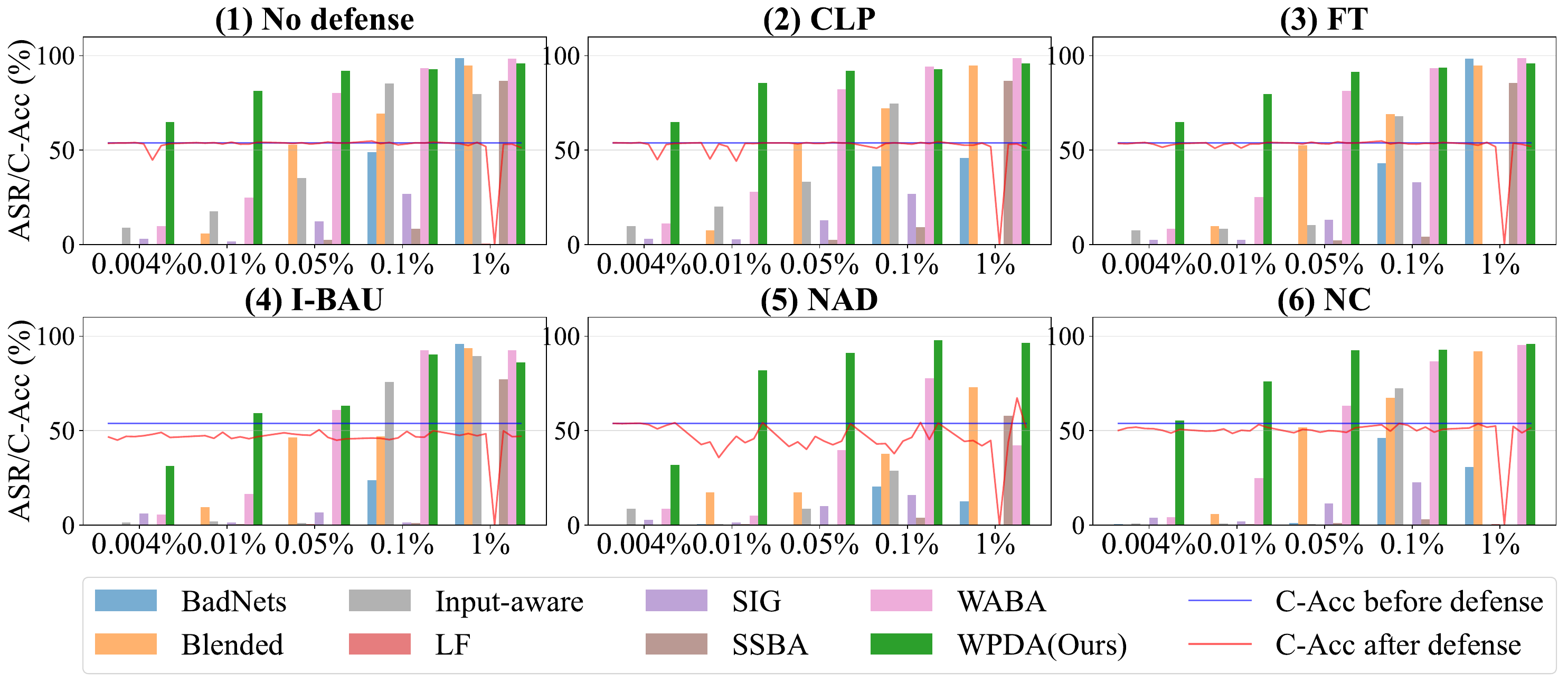}
    \label{tiny_vgg_defense}}
\caption{Results of 7 compared attack methods and WPDA against 5 defense methods on VGG19-BN, including CLP, FT, I-BAU, NAD and NC. Each sub-plot corresponds to one defense method. In each sub-plot, each bar in the histogram indicates the ASR value of one attack method, and the C-Acc values are represented by curves.}
\vspace{-0.5cm}
\label{all-defense-vgg}
\end{figure}

In the following, we evaluate the resistance performance of 7 compared attack methods and WPDA under 5 SOTA backdoor defense methods on VGG19-BN. The results on PreAct-ResNet18 are shown in Supplementary
% ~\ref{defense on pre}.
\paragraph{\textbf{Evaluations on CIFAR-10}}
% \textbf{Evaluations on CIFAR-10}.
Fig.~\ref{CIFAR-10_vgg_defense} illustrates the performance of 7 SOTA attack methods and WPDA against 5 SOTA defense methods. Since the compared methods fail to achieve backdoor attacks at low poisoning ratios, particularly at the extremely low poisoning ratio of 0.004\%, their resistance are impossible to be evaluated. In contrast, WPDA achieves successful backdoor attacks across all poisoning ratios and retains effective attack performance after undergoing defenses.
% Compared with other attack methods, WPDA shows resistant performance after various defense methods at every poisoning ratio. 
% Besides, backdoor attacks at low poisoning ratios are more resistant than those at high poisoning ratios in some cases. For instance, the ASR of attack methods, such as Blended, Input-aware and WPDA at 0.05\% poisoning ratio, decays less than that at 0.1\% poisoning ratios after 5 defense methods. 

\paragraph{\textbf{Evaluations on CIFAR-100}}
Fig.~\ref{CIFAR-100_vgg_defense} illustrates the performance of 7 compared attack methods and WPDA against defenses. At 1\% poisoning ratio, some attack methods illustrate a significant decline in attack efficiency after undergoing defenses. For instance, after undergoing NC, the ASR of BadNets, Blended, and Input-aware drops substantially. However, after undergoing CLP, although the attacks of Input-aware and WPDA become ineffective, the C-Acc of the model is significantly damaged, indicating that CLP struggles to defend the attacks from Input-aware and WPDA. At the extremely low poisoning ratio of 0.004\%, the attack performance of WPDA is weakened after CLP, but the ASR remains. Overall, WPDA demonstrates resistance even after 5 SOTA defenses, and outperforms 7 compared methods, particularly at extremely low poisoning ratios.

\paragraph{\textbf{Evaluations on Tiny ImageNet}}
Fig.~\ref{tiny_vgg_defense} illustrates the excellent performance of WPDA on Tiny ImageNet after undergoing 5 different defenses. At 0.004\% poisoning ratio, despite a decrease in ASR after I-BAU and NAD, WPDA retains its optimal attack effectiveness. At 1\% poisoning ratio, Input-aware becomes ineffective after undergoing CLP, FT, NAD and NC, as well as SSBA after NC. These results confirm that lots of backdoor attacks based on high poisoning ratios are at risk of being defended. WPDA withstood 5 defenses across all poisoning ratios, exhibiting great resistance and demonstrating superiority over 7 compared attacks, particularly at extremely low poisoning ratios.
\paragraph{\textbf{Summary}} 

Due to inherent variations among datasets, the performance of a single defense method can vary across different datasets. However, as the results indicate, most defense methods struggle to effectively defend backdoor attack methods at low poisoning ratios, 
while WPDA successfully leverages this characteristic to withstand these defenses.  Compared to 7 SOTA attack methods, WPDA exhibits strong resistance.
% underscoring the necessity of developing defense methods specifically designed for low poisoning ratios. Compared to 7 SOTA attack methods, WPDA demonstrates more resistant, stealthy and effective performance, especially at low poisoning ratios.

% Most defense methods are difficult to defend against backdoor attacks with low poisoning ratios, which indicates that it is necessary to explore backdoor attacks at low poisoning ratios. Compared to 7 SOTA attack method, WPDA shows more resistant and stealthy performance, especially at low poisoning ratios.
%%%%%%%%%%%%%%%%%%%%%%%%%%%%%%%%%%%%%%%%%%%%%%%%%%%%%%%%%%%%%%%%%%%%%%%%%%%%%%%%%%%%%%%%%%%%%%%%%%%%%%%%%%%%%%%%%%%%%
\begin{figure}[htbp]
\centering
\includegraphics[width=3.2in]{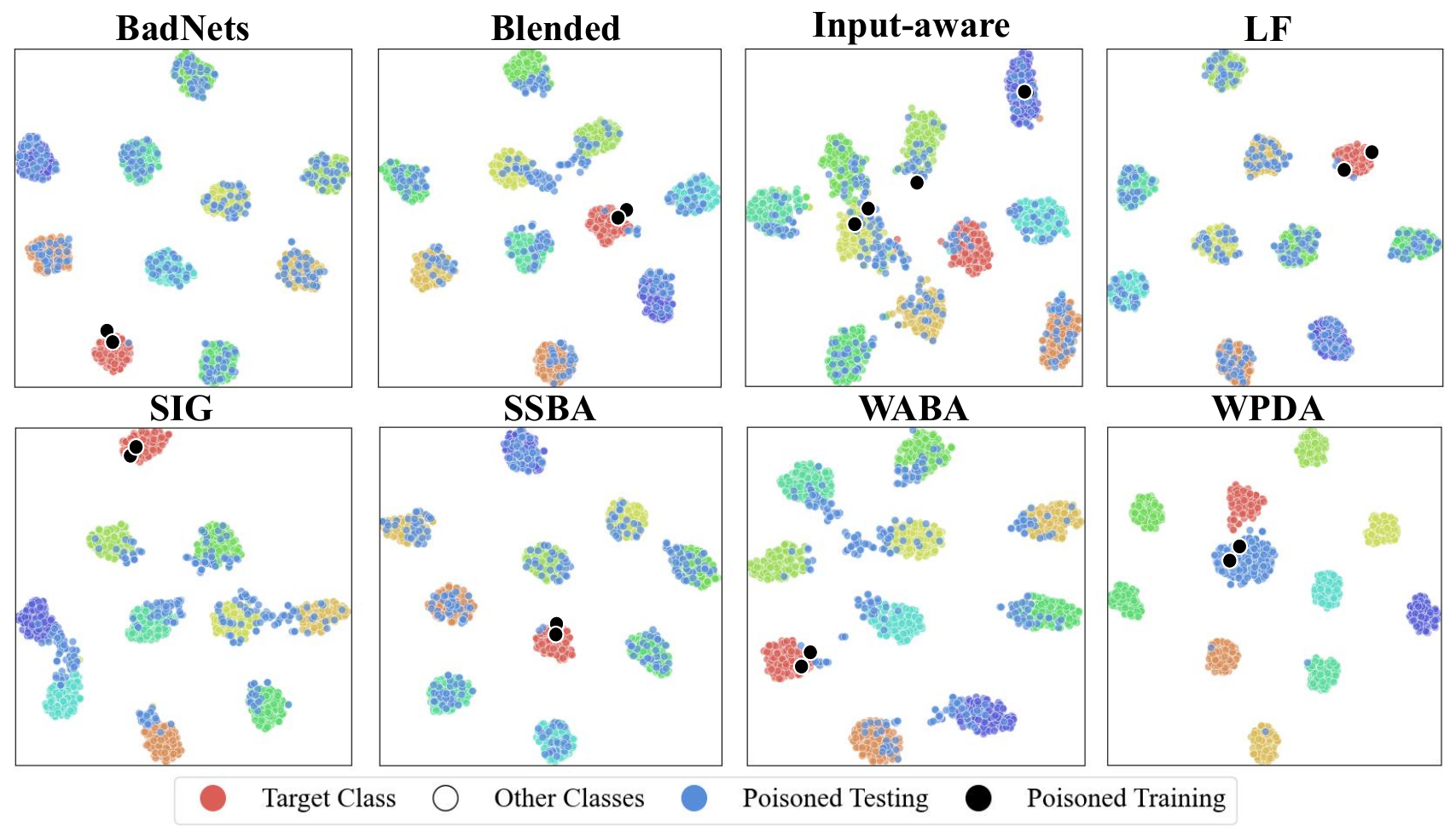}
\caption{t-SNE\cite{van2008visualizing} about 7 SOTA attack methods and WPDA at 0.004\% poisoning ratio.}
\label{tsne-00004-all-attack}
\vspace{-0.5cm}
\end{figure}

\vspace{-0.2cm}
\section{Analyses}
\label{tsne-kde}
\subsection{Visualization on t-SNE}
Fig.~\ref{tsne-00004-all-attack} illustrates the t-SNE of 7 SOTA attacks and WPDA on PreAct-ResNet18 trained by CIFAR-10 at 0.004\% poisoning ratio. The red circles indicate benign training samples of the target class, the black circles indicate poisoned training samples and the blue circles indicate poisoned testing samples, both of which are labeled as the target class. For BadNets, Blended, Input-aware, SIG, SSBA, LF and WABA, it is evident that the poisoned training samples are close to the benign samples in the target class, while the poisoned testing samples are evenly distributed across the classes. Based on this observation, we can infer that 
the models based on these attack methods only memorise the poisoned training samples rather than learning the features of the trigger, which results in the backdoor not being activated when the poisoned testing samples are used as inputs. 
Regarding WPDA, the poisoned training samples, the benign testing samples in the target class and the poisoned testing samples are relatively clustered in the feature space, indicating that the features extracted by the model from those samples exhibit significant similarity and the model effectively learns the features of the trigger. As a result, WPDA precisely inserts trigger information into the critical frequency regions, promoting the model to effectively learn the trigger information even at low poisoning ratios. 
\vspace{-0.2cm}
\subsection{{Visualization on \texorpdfstring{$l_2$}--distance KDE}}
\begin{figure}[!ht]
\centering  %图片全局居中
% \captionsetup[subfigure]{labelformat=empty} 
    {
    \includegraphics[width=1.70in]{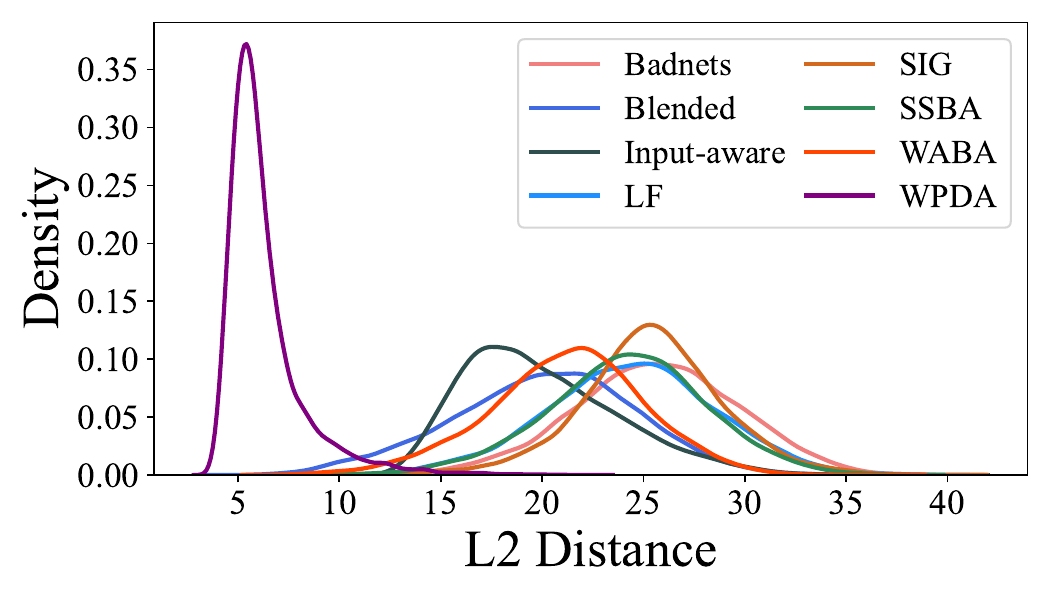}}
    \label{kde-00004}
% \captionsetup[subfigure]{labelformat=empty} 
\hspace{-0.3cm}
    {
    \includegraphics[width=1.70in]{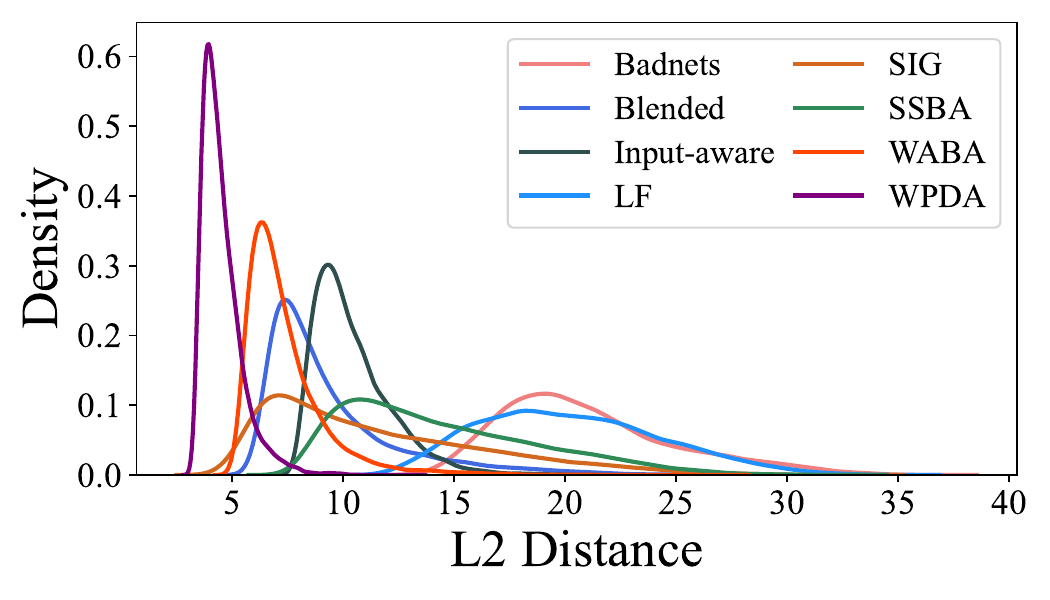}}
    \label{kde-01}
\caption{$l_2$-distance KDE on CIFAR-10 at 0.004\% poisoning ratio (left) and 1\% poisoning ratio (right).}
\label{kde}
\vspace{-0.1cm}
\end{figure}

% Kernel Density Estimation (KDE) is a non-parametric statistical method for estimating the Probability Density Function (PDF) that can be used to describe the distribution of data~\cite{gramacki2018nonparametric}. Specifically, KDE is a measure of estimated probability density defined by a kernel function near each data point in the dataset. In KDE, we usually choose a kernel function and a bandwidth parameter to define the similarity between each data point in the dataset and other data points, usually using the Gaussian kernel function, \etc~\\
\hspace*{0.15cm}
To further verify the effectiveness of the WPDA attack, we conduct $l_2$-distance visualization analyses based on KDE~\cite{gramacki2018nonparametric}
% on PreAct-ResNet18 trained by CIFAR-10 
at different poisoning ratios. We extract the features of poisoned samples from the `linear' layer in PreAct-ResNet18 and calculate the $l_2$-distance between each poisoned training sample and each poisoned testing sample, then we can obtain multiple lists of $l_2$-distance between each poisoned training sample and all poisoned testing samples, where the number of lists is equal to the number of poisoned training samples and the length of each list is equal to the number of poisoned testing samples. 
We then average the multiple lists of $l_2$-distance and conduct KDE visualization. 
Fig.~\ref{kde} illustrates $l_2$-distance KDE on PreAct-ResNet18 trained by CIFAR-10. The horizontal axis represents the average $l_2$-distance between the poisoned training samples and the poisoned testing samples, and the vertical axis represents the density of each $l_2$-distance value. The $l_2$-distance with a large value means that the features extracted by the model from the poisoned training samples have a significant variability from the features extracted from the poisoned testing samples. It is much difficult for the model to predict the poisoned testing samples as target class.
At 0.004\% poisoning ratio, the $l_2$-distance values in WPDA are mainly concentrated in small values, which indicates that the features extracted by the model from the poisoned testing samples are similar to those extracted in the poisoned training samples. In contrast, for other attacks, the features extracted by the model from the poisoned testing samples differ from those extracted from the poisoned training samples at 0.004\% poisoning ratio. As the poisoning ratios increase, leading to a rise in the ASR, $l_2$-distance values in each attack decrease and gradually concentrate, indicating that the model is progressively learning the trigger information.
% \vspace{-0.3cm}

%%%%%%%%%%%%%%%%%%%%%%%%%%%

\begin{figure}[!ht]
\centering  %图片全局居中
% \captionsetup[subfigure]{labelformat=empty} 
    {
    \includegraphics[width=1.70in]{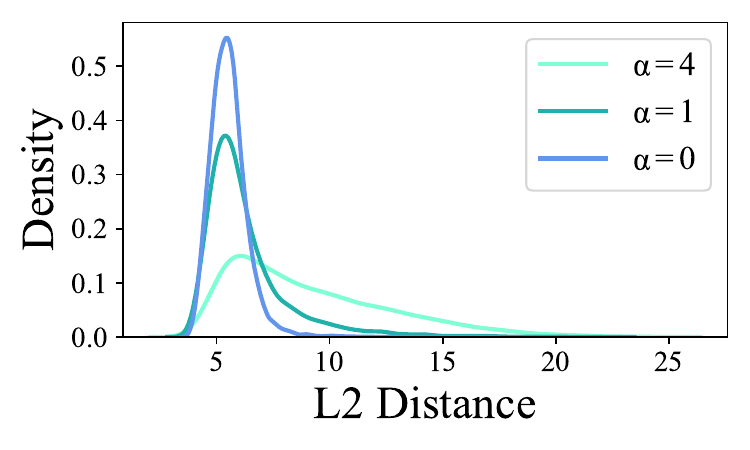}}
    \label{kde-alpha}
% \captionsetup[subfigure]{labelformat=empty} 
\hspace{-0.3cm}
    {
    \includegraphics[width=1.70in]{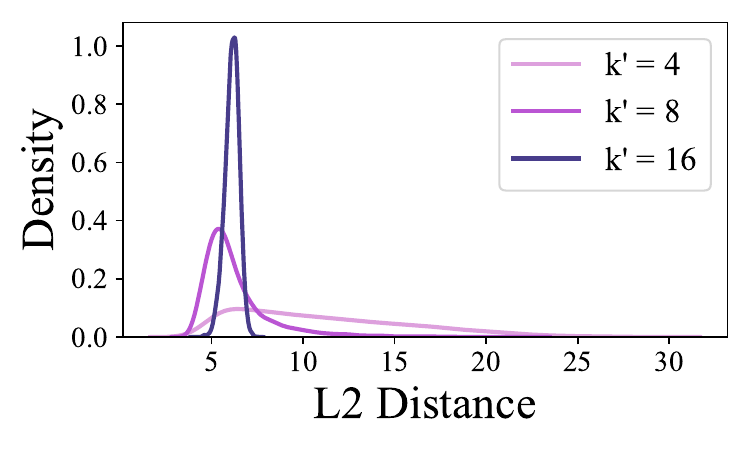}}
    \label{kde-k'}
\caption{$l_2$-distance KDE on CIFAR-10 at 0.004\% poisoning ratio across different $\alpha$ (left) and different $k'$ (right).}
\label{kde-generation}
\vspace{-0.1cm}
\end{figure}

Fig.~\ref{fig:generation} has demonstrated that trigger exhibits generalization capability. To further confirm our conclusion, we conduct $l_2$-distance KDE visualization on PreAct-ResNet18 by modifying original images information intensity $\alpha$ and trigger information intensity $k'$ of the poisoned testing samples.
% Fig.~\ref{kde-alpha} 
Fig.~\ref{kde-generation} shows the $l_2$-distance KDE visualization under different original images information intensities $\alpha$. With the increase of the intensity of original images information $\alpha$, the $l_2$-distance in feature space between the poisoned testing samples and the poisoned training samples gradually increases. As a result, the original images information has a negative effect on the backdoor activation, but we can rely on the models' generalization ability to activate the backdoor when the prominence of the trigger is stronger than that of the original images information.
% Fig.~\ref{kde-k'} 
Fig.~\ref{kde-generation} shows the $l_2$-distance KDE visualization under different trigger information intensities $k'$. The increase of the trigger information intensity $k'$ can promote the backdoor model to capture the trigger information, which is positive for the activation of the backdoor.
\vspace{-0.5cm}
% \begin{figure}[!ht]
% \centering  %图片全局居中
%     \subfigure[$l_2$-distance KDE under different original benign sample information intensities $\alpha$]{
%     \label{kde-alpha}
%     \includegraphics[width=1.7in]{figures/tsne_WPDA_alpha_00004_.pdf}}
%     \subfigure[$l_2$-distance KDE under different trigger information intensities $k'$]{
%     \label{kde-k'}
%     \includegraphics[width=1.7in]{figures/tsne_WPDA_k1_00004_.pdf}}
% \caption{$l_2$-distance KDE on CIFAR-10 at 0.004\% poisoning ratio.}
% \label{kde-generation}
% \end{figure}

\vspace{0.2cm}
\section{Conclusion}
\label{conclusion}
In this work, a novel frequency-based dataset-specific backdoor attack, WPDA, is proposed. By analyzing the frequency information distribution of datasets based on Wavelet Packet Decomposition (WPD), we infer the critical frequency regions which the models would focus on and insert trigger. To achieve stealthiness, effectiveness, and resistance, we provide characteristic analyses experiments to design backdoor pattern.
Our method outperforms other existing SOTA backdoor attack methods, especially at extremely low poisoning ratios, and bypasses the advanced backdoor defenses.
% Our method not only maintain the naturalness of backdoor samples, but also successfully achieve backdoor attack at extremely low poisoning ratios. 
% Besides, we conduct $l_2$-distance KDE visualization experiments to observe the $l_2$-distance value of attack methods at different poisoning ratios, demonstrating that WPDA promotes the model to efficiently learn the trigger information rather than memorizing the poisoned training samples at low poisoning ratios.
Besides, we conduct t-SNE to analyze the distribution of the poisoned training samples and the poisoned testing samples in the feature space to vertify the effectiveness of our method. We also conduct $l_2$-distance KDE visualization experiments to observe the $l_2$-distance value of attack methods at different poisoning ratios, demonstrating that WPDA promotes the model to efficiently learn the trigger information rather than memorising the poisoned training samples at low poisoning ratios.
% \vspace{-0.3cm}
\section*{Acknowledgments}
The authors would like to thank National Natural Science Foundation of China under grant No.62076213, Shenzhen Science and Technology Program under grants No.RCYX20210609103057050, which support Baoyuan Wu.
\bibliography{ref_tip_new}
\bibliographystyle{IEEEtran}
% \newpage
% \newpage

% \input{tip/7_appendix_new2}
% \input{tdsc/8_supporting_document}
% \newpage

 

% \end{IEEEbiography}



% \vfill

\end{document}